\documentclass[12pt]{article}

\usepackage[totalheight = 23cm, totalwidth = 17cm]{geometry}
\usepackage{amssymb,amsmath,amsfonts,amsbsy,graphicx,bm}
\usepackage{color}

\newcommand{\mpl}{m_{\rm Pl}}
\newcommand{\calA}{{\cal A}}
\newcommand{\calC}{{\cal C}}

\newcommand{\calE}{{\cal E}}
\newcommand{\calH}{{\cal H}}
\newcommand{\calL}{{\cal L}}
\newcommand{\calO}{{\cal O}}
\newcommand{\calR}{{\cal R}}
\newcommand{\calS}{{\cal S}}

\renewcommand{\theequation}{\arabic{section}.\arabic{equation}}

\begin{document}

\begin{titlepage}

\begin{center}

{\LARGE \bf 
Quantum non-linear evolution of \\ inflationary tensor perturbations
}

\vskip 1.0cm

{\large
Jinn-Ouk Gong$^{a}$ 
and Min-Seok Seo$^{b}$ 
}

\vskip 0.5cm

{\it
$^{a}$Korea Astronomy and Space Science Institute, Daejeon 34055, Korea
\\
$^{b}$Department of Physics Education, Korea National University of Education
\\ 
Cheongju 28173, Korea
}

\vskip 1.2cm

\end{center}

\begin{abstract}

We study the quantum mechanical evolution of the tensor perturbations during inflation with non-linear tensor interactions. We first obtain the Lindblad terms generated by non-linear interactions by tracing out unobservable sub-horizon modes. Then we calculate explicitly the reduced density matrix for the super-horizon modes, and show that the probability of maintaining the unitarity of the squeezed state decreases in time. The decreased probability is transferred to other elements of the reduced density matrix including off-diagonal ones, so the evolution of the reduced density matrix describes the quantum-to-classical transition of the tensor perturbations. This is different from the classicality accomplished by the squeezed state, the suppression of the non-commutative effect, which is originated from the quadratic, linear interaction, and also maintains the unitarity. The quantum-to-classical transition occurs within 5 - 10 $e$-folds, faster than the curvature perturbation.

\end{abstract}

\end{titlepage}

\newpage

\section{Introduction}
\setcounter{equation}{0}

While inflation~\cite{Guth:1980zm,Linde:1981mu,Albrecht:1982wi} is postulated to resolve initial condition problems of the early universe, it also gives a natural account of how to generate the initial perturbations on super-horizon scales as observed in the cosmic microwave background (CMB)~\cite{Akrami:2018odb}. That is, during inflation the primordial quantum fluctuations leave the horizon and become classical perturbations as we observe today such as the temperature anisotropies in the CMB and inhomogeneously distributed galaxies~\cite{Mukhanov:1981xt,Guth:1982ec,Hawking:1982cz,Starobinsky:1982ee,Bardeen:1983qw}.  As for this aspect, the quantum-to-classical transition is known to arise as the states for the scalar and tensor perturbations take a form of specific coherent superposition of excitations, called the squeezed state~\cite{Guth:1985ya}. More concretely, through time evolution, some quadratic terms reflecting curved space-time make the state ``squeezed'' (see Appendix~\ref{app:quadratic}) such that the non-commutative effect between the perturbation field variable and its canonical conjugate momentum becomes negligibly small. The probability density of the squeezed state is not localized in both the field variable and its canonical momentum, but spreads over the phase space. Instead, the suppression of the quantum mechanical nature of non-commutativity allows us to interpret the probability distribution as a classical one, just like the semi-classicality of the Wentzel-Kramers-Brillouin (WKB) approximation. Such a squeezing of the state has been discussed in various contexts in physics (see, e.g.~\cite{Grishchuk:1992tw,Albrecht:1993hf}). As an application to the inflationary cosmology, generic massless particle case was discussed~\cite{Polarski:1995jg} and analyzed in terms of tensor~\cite{Grishchuk:1992tw,Grishchuk:1990bj} and curvature~\cite{Albrecht:1992kf} perturbations, respectively.

On the other hand, the presence of the horizon in (quasi) de Sitter (dS) space-time during inflation enables us to separate the quantum fluctuation modes into those shorter and longer than the horizon scale according to their wavelengths. When we focus on the modes relevant for observations, say, the modes with wavelengths longer than the horizon scale, they are evolving non-linearly in the time-dependent background due to gravity, interacting with sub-horizon modes which are not very relevant observationally. In that sense, the system of the super-horizon modes is open. After tracing out unobservable modes, or environment, the non-linear interactions between modes in system and environment introduce the so-called Lindblad operators~\cite{Lindblad:1975ef} which violate the unitary evolution of the density matrix. Since the Lindblad terms are equivalent to adding new Hamiltonian terms with randomly varying source~\cite{Banks:1983by}, they have a close connection to the stochastic inflation formalism~\cite{Starobinsky:1986fx}. Indeed, inclusion of the Lindblad terms provide the most generic evolution of the density matrix under several assumptions: the evolution equation of the density matrix is Markovian, i.e. the density matrix at some specific time depends only on that at an earlier time, rather than that over a range of times, and linear in the density matrix. Moreover, the resulting density matrix is Hermitian, and has a unit trace with a positive expectation value, satisfying probability interpretation for the ``mixed state''~\cite{Pearle:2012}.

It has been known that the time evolution regulated by the Lindblad opeartors leads to the loss of quantum nature during inflation in another way. One way to see this is to investigate the ``decoherence'' -- the decay of the interference effect~\cite{Dowker:1992es,GellMann:1992kh,Gell-Mann:2018dzd}, mimicking collapse of the wave function into some specific eigenstate of the observable (for a review on decoherence in generic context, see e.g.~\cite{Zurek:2003zz,Schlosshauer:2003zy}). Recent studies~\cite{Burgess:2006jn,Burgess:2014eoa,Nelson:2016kjm,Shandera:2017qkg,Martin:2018zbe} on the Lindblad operators during inflation emphasized this aspect. Meanwhile, the Lindblad operators also reveal the classicality of the system by converting the pure state -- for inflation, the squeezed state -- into the mixed state. While the Lindblad operators introduce non-unitary evolution of the pure state, they still maintain the trace of the density matrix unity. Then the pure state evolves into the mixed state, giving a non-trivial density matrix elements representing classical probability. This indeed invites us to use the Lindblad operators to describe the information paradox in Hawking radiation~\cite{Banks:1983by} and also similar phenomena in the inflationary cosmology.

Therefore, we have two different aspects for the quantum-to-classical transition. First, the linear evolution leading to the squeezed state is evidently unitary process. Moreover, it corresponds to the quadratic sector in the action, in which the system-environment interaction does not appear. On the contrary, the Lindblad operators lead to non-unitary time evolution in both decoherence and generation of the mixed state. It comes from the system-environment interaction, which appears through non-linear interactions beyond quadratic order. Tiny non-Gaussianity constrained by recent observations on the CMB~\cite{Ade:2015ava} implies that the non-linear effect is sub-dominant. Then we expect that the WKB-like ``unitary'' classicality achieved by the squeezed state, often referred to as the semi-classicality, is slowly converted into the ``non-unitary'' classicality through the Lindblad operators as a perturbation. The Lindblad operators provide the rate of such conversion, reflecting the specific form of the non-linear interactions.

Previous studies on the non-unitary classicality considered the curvature perturbation as an example~\cite{Burgess:2006jn, Burgess:2014eoa, Nelson:2016kjm, Shandera:2017qkg}. The curvature perturbation is interpreted as a Goldstone boson resulting from the spontaneous breaking of dS isometry by quasi dS background, which is parametrized by the slow-roll parameters~\cite{Cheung:2007st} (see also \cite{Prokopec:2010be,Gong:2016qpq}). This implies that small slow-roll parameters make the conversion rate of unitary to non-unitary classicality very small, as analyzed in e.g.~\cite{Nelson:2016kjm}. For the tensor perturbations, on the other hand, situation is different. Unlike the curvature perturbation, they are well defined even in perfect dS space-time, so the leading effect is irrelevant to the slow-roll parameters.

The purpose of this article is to explore the effect of the Lindblad operators in the evolution of pure tensor perturbations. Especially, we concretely study the evolution of the pure squeezed state into mixed state to show how unitary classicality is converted into non-unitary classicality by obtaining the conversion rate. In Section~\ref{sec:reduced}, we describe how the Lindblad operators are given when tracing out environment. In Section~\ref{sec:Hint}, we consider the cubic interaction of the tensor perturbations from which we obtain the Lindblad operators explicitly in Section~\ref{sec:structure}. This provides a non-unitary evolution equation of the density matrix. By applying this to the evolution of the squeezed state, we calculate the conversion rate of unitary classicality  into non-unitary classicality for the tensor perturbation in Section~\ref{sec:solution}. A brief conclusion is following in Section~\ref{sec:conclusion}. In Appendix~\ref{app:quadratic}, we review how the time evolution generated by quadratic sector leads to squeezed state, with our convention for notation. Details of calculations throughout our work are sketched in the other appendix sections.

\section{Non-linear evolution of reduced density matrix}
\label{sec:reduced}
\setcounter{equation}{0}

We begin with a general account on the evolution of a system where we have integrated out certain sector. Given that initially the whole system is composed of the ``fast'' environment, or bath, and relatively ``slow'' system of our interest, whose state vectors respectively denoted by $|\calE(\tau_0)\rangle$ and $|\calS(\tau_0)\rangle$ so that the initial state vector of the whole system is written as $|\Psi\rangle = |\calS(\tau_0)\rangle|\calE(\tau_0)\rangle$, the density matrix, or density operator, $\rho$ is given by
\begin{equation}
\rho(\tau_0) \equiv |\Psi\rangle\langle\Psi| 
= |\calE(\tau_0)\rangle|\calS(\tau_0)\rangle\langle\calS(\tau_0)|\langle\calE(\tau_0)| 
\, .
\end{equation}
The evolution of the density operator, with interactions, can be written in a different manner in which picture one is using. One general remark however is that the density matrix is not an usual operator, in the sense that it follows the von Neumann equation
\begin{equation}
\frac{d\rho}{d\tau} = -i \big[ H, \rho \big] \, ,
\end{equation}
where the sign is opposite to the standard Heisenberg equation.

We basically work in the Schr\"odinger picture. The reason is, despite of the existence of higher-order interactions, the evolution of the whole system is contained in the states in the Schr\"odinger picture, from which we construct the density matrix of our interest. Since in general the Hamiltonian is time-dependent and not commuting at different times, the density matrix in the Schr\"odinger picture is given by, up to second order explicitly in interaction,
\begin{align}
\label{eq:sol-s.pic}
\rho(\tau) 
& = 
\rho_\text{free}(\tau) 
- i U_0(\tau;\tau_0)
\int_{\tau_0}^\tau d\tau_1 \Big[ H_{\text{int},I}(\tau_1), \rho(\tau_0) \Big]
U_0^\dag(\tau;\tau_0)
\nonumber\\
& \quad
+ (-i)^2 U_0(\tau;\tau_0)
\int_{\tau_0}^\tau d\tau_2 \int_{\tau_0}^{\tau_2} d\tau_1
\Big[ H_{\text{int},I}(\tau_2) \Big[ H_{\text{int},I}(\tau_1), \rho(\tau_0) \Big] \Big] 
U_0^\dag(\tau;\tau_0)
+ \cdots
\nonumber\\
& \equiv 
\rho_\text{free}(\tau) + \rho_{(1)}(\tau) + \rho_{(2)}(\tau) 
+ \cdots 
\, ,
\end{align}
where 
\begin{equation}
U_0(\tau;\tau_0) \equiv T e^{-i\int_{\tau_0}^\tau H_0(\tau') d\tau'} 
\end{equation}
is the free evolution operator with $T$ ($\overline{T}$) denoting (anti-) time-ordering operator, 
\begin{equation}
\rho_\text{free}(\tau) 
= U_0(\tau;\tau_0) \rho(\tau_0) U_0^\dag(\tau;\tau_0)
\end{equation}
is the ``free'' density matrix in the Schr\"odinger picture at $\tau$ in the absence of interaction, and $H_{\text{int},I}(\tau)$ is the interaction Hamiltonian in the interaction picture. We are eventually interested in the ``reduced'' density matrix where the effects of the fast modes, or the environment, are integrated out. So we take the trace over the environment state $|\calE_i\rangle$ at the time $\tau$, {\it not} the state $|\calE(\tau)\rangle = U_0(\tau;\tau_0)|\calE(\tau_0)\rangle$ which has evolved from the initial environment state $|\calE(\tau_0)\rangle$, with $i$ being an abstract index for the a possible choice of the basis states for environment
at $\tau$:
\begin{equation}
\label{eq:rho_red}
\rho_\text{red}(\tau) 
= 
\sum_i \langle\calE_i| \rho(\tau) |\calE_i\rangle 
= 
\text{Tr}_\calE \rho \, .
\end{equation}
Being a choice of the basis states, $|\calE_i\rangle$ is complete:
\begin{equation}
\label{eq:complete}
\sum_i |\calE_i\rangle \langle\calE_i| = 1 
\, .
\end{equation}

The time evolution of this reduced density matrix can be found as:
\begin{align}
\frac{d\rho_\text{red}}{d\tau} 
= 
\frac{d}{d\tau} \sum_i 
\langle\calE_i| \rho(\tau) |\calE_i\rangle
= 
\sum_i \bigg\langle\calE_i\bigg| \frac{d\rho_S}{d\tau} \bigg|\calE_i\bigg\rangle
= 
\text{Tr}_\calE \frac{d\rho}{d\tau} \, ,
\end{align}
since the environmental states we trace out are {\it fixed} at the time of tracing out the environment, $|\calE_i\rangle$ does not carry time-dependence. Thus to obtain the evolution equation for the reduced density matrix, we only need to take the trace of that for the full density matrix (however, see the discussion below). To proceed, we note that for $H_0$ we can completely separate the system and environment since we are considering free evolution, so $H_0 = H_{0,\calS} + H_{0,\calE}$ with $\big[ H_{0,\calS}, H_{0,\calE} \big] = 0$. To trace over $|\calE_i\rangle$, we note that since $|\calE_i\rangle$ is the environment state at a later time $\tau$, it has no overlap with the initial system state $|\calS(\tau_0)\rangle$, i.e. $|\calE_i\rangle$ and $|\calS(\tau_0)\rangle$ are orthogonal. Meanwhile, the initial environment state $|\calE(\tau_0)\rangle$ may still have ample overlap with $|\calE_i\rangle$, so only $|\calE(\tau_0)\rangle$ responds to $|\calE_i\rangle$. Then, after some calculations, we find the evolution equation for $\rho_\text{red}$ as (see~\cite{Shandera:2017qkg,Martin:2018zbe} for derivation)
\begin{align}
\label{eq:Lindblad}
\frac{d}{d\tau}\rho_\text{red}(\tau) 
& =
-i \Big[ H_{0,\calS}, \rho^{(0)}_\text{red} + \rho^{(1)}_\text{red} 
+ \rho^{(2)}_\text{red} \Big]
-i \Big[ H_{\text{int}}^\text{(eff1)} + H_{\text{int}}^\text{(eff2)}, 
\rho_\calS(\tau) \Big]
\nonumber\\
& \quad
- \frac{1}{2} \sum_i \Big[
L_1^\dag L_2 \rho_\calS(\tau) + \rho_\calS(\tau)L_2^\dag L_1
- 2L_1\rho_\calS(\tau)L_2^\dag + (L_1 \leftrightarrow L_2) 
\Big] \, .
\end{align}

\begin{itemize}

\item With
\begin{equation}
\label{eq:rho-freeS}
\rho_\calS(\tau)
\equiv 
U_{0,\calS}(\tau;\tau_0) |\calS(\tau_0)\rangle \langle\calS(\tau_0)|
U^\dag_{0,\calS}(\tau;\tau_0)
\, ,
\end{equation}
where $U_{0,\calS}(\tau;\tau_0) = T e^{-i\int_{\tau_0}^\tau H_{0,\calS}}$, being the free-evolved density matrix of the system states, in the first term in \eqref{eq:Lindblad}, we have defined
\begin{equation}
\label{eq:rho0red-initial}
\rho^{(0)}_\text{red}(\tau) 
\equiv 
\rho_{\calS}(\tau) 
\sum_i \bigg| \Big\langle\calE_i\Big| U_{0,\calE}(\tau;\tau_0)  
\Big|\calE\Big\rangle \bigg|^2 
\, ,
\end{equation}
where $U_{0,\calE}(\tau;\tau_0) = T e^{-i\int_{\tau_0}^\tau H_{0,\calE}d\tau'}$ (so that $U_0 = U_{0,\calS}U_{0,\calE}$), $\rho^{(1)}_\text{red}(\tau) \equiv \sum_i \langle\calE_i|\rho_{(1)}(\tau)|\calE_i\rangle$ and $\rho^{(2)}_\text{red}(\tau) \equiv \sum_i \langle\calE_i|\rho_{(2)}(\tau)|\calE_i\rangle$. We can write this term purely in terms of the commutators with free Hamiltonian of the system so the evolution of the reduced density matrix described by this term is unitary. Now, with time-dependent background there is ambiguity how to choose the system and environment states. We may take the state $|\calE_i\rangle$ to be traced out as the environmental Fock space, i.e. the modes on sub-horizon scales {\it fixed} at the moment $\tau$, viz. those with $k > aH = -1/\tau$. This is quite natural, but there is no obvious sharp distinction between the initial system and environment states $|\calS(\tau_0)\rangle$ and $|\calE(\tau_0)\rangle$. Different choices could well be justified for different reasons. One choice that is at least computationally appealing is such that we choose $|\calE(\tau_0)\rangle$ in such a way that this corresponds to the modes up to which becomes the horizon scale $k = aH = -1/\tau$ at the moment $\tau$, and $|\calS(\tau_0)\rangle$ denotes all the modes with smaller $k$. Then with \eqref{eq:complete} and $\langle \calE(\tau_0)|\calE(\tau_0)\rangle=1$, some expressions become very simple, e.g. $\rho^{(0)}_\text{red} = \rho_{\calS}$ in \eqref{eq:rho0red-initial}.

\item In the second term in \eqref{eq:Lindblad}, we have defined
\begin{align}
H_{\text{int}}^\text{(eff1)}
& \equiv
\Big\langle \calE(\tau_0) \Big| U^\dag_{0,\calE}(\tau;\tau_0)
H_{\text{int},S} U_{0,\calE}(\tau;\tau_0) \Big| \calE(\tau_0) \Big\rangle 
\, ,
\nonumber\\
H_\text{int}^\text{(eff2)}
& \equiv 
\frac{-i}{2} \sum_i \Big( L_1^\dag L_2 
- L_2^\dag L_1 \Big)
\, ,
\end{align}
where $H_{\text{int},S}$ is the interaction Hamiltonian in the Schor\"odinger picture and $L_i$'s are the so-called Lindblad operators defined below. While written purely in terms of the commutator form, one thing we can further note is the existence of the ``shift'' in the Hamiltonian, $H_\text{int}^\text{(eff1)} + H_\text{int}^\text{(eff2)}$. This means even the unitary part of the evolution is described by a Hamiltonian not identical to the unperturbed free Hamiltonian of the system, because the environment perturbs the free Hamiltonian in such a way that the free Hamiltonian that describes the unitary evolution should be in general shifted. Note that as can be inferred, $H_\text{int}^\text{(eff2)}$ and the Lindblad terms given below originate from the same term, $-i \sum_i \Big\langle \calE_i \Big| \Big[ H_{\text{int},S}, \rho_{(1)}(\tau) \Big] \Big| \calE_i \Big\rangle$.

\item Lastly, in the third term in \eqref{eq:Lindblad} we have defined the Lindblad operators as
\begin{align}
\label{eq:Lindblad-op1}
L_1 & \equiv \Big\langle \calE_i \Big| H_{\text{int},S}
U_{0,\calE}(\tau;\tau_0) \Big|\calE(\tau_0)\Big\rangle \, ,
\\
\label{eq:Lindblad-op2}
L_2 & \equiv \bigg\langle\calE_i\bigg|
\int_{\tau_0}^\tau d\tau_1 H_{\text{int},I}(\tau_1-\tau)
U_{0,\calE}(\tau;\tau_0) \bigg|\calE(\tau_0)\bigg\rangle \, .
\end{align}
They come from the pure commutator contributions between the density matrix and the interaction Hamiltonian, and are responsible for the non-unitary evolution of the reduced density matrix: since the third term in \eqref{eq:Lindblad} is real, this part does not lead to a simple sinusoidal oscillations but gives rise to an exponential change in some elements of the reduced density matrix. Or, some information on the environment is integrated out, now the whole system is in the so-called mixed state, or decoherence occurred. This is described in several articles for the scalar cosmological perturbation, i.e. the curvature perturbation  \cite{Burgess:2006jn, Burgess:2014eoa, Nelson:2016kjm, Shandera:2017qkg, Martin:2018zbe}. Also note that the trace of the Lindblad terms is obviously zero, meaning that the total classical probability is preserved. This leads to interesting consequences as we will see in Section~\ref{sec:solution}.

\end{itemize}

Before closing, it is illustrative to consider what if the contributions are only from the system sector. Then, the environment state $|\calE_i\rangle$ commutes with $H_\text{int}$, so \eqref{eq:Lindblad-op1} and \eqref{eq:Lindblad-op2} become 
\begin{align}
L_1 & = H_{\text{int},S} \, ,
\\
L_2 & = \int_{\tau_0}^\tau d\tau_1 H_{\text{int},I}(\tau_1-\tau) \, .
\end{align}
Then we can easily find
\begin{align}
\label{eq:Hint-onlysystem}
& -i \sum_i \Big\langle \calE_i \Big| \Big[ H_{\text{int},S}, \rho_{(1)}(\tau) \Big] \Big| \calE_i \Big\rangle
= 
-i \bigg[ H_{\text{int},S}, -i U_{0,\calS}(\tau;\tau_0)
\int_{\tau_0}^\tau d\tau_1 \Big[ H_{\text{int},I}(\tau_1),\rho_{\calS}(\tau_0) \Big]
U^\dag_{0,\calS}(\tau;\tau_0)
\bigg] 
\, ,
\end{align}
where $\rho_{\calS}(\tau_0) \equiv \big|\calS(\tau_0)\big\rangle \big\langle\calS(\tau_0)\big|$. What happens is, not surprisingly, that we now have the unitary evolution of the pure system part in the precisely equivalent form to the standard von Neumann equation. That means, if we consider the contribution that purely belongs to the system sector, not including any from the environment sector, such a contribution can be absorbed into the unitary evolution, not contributing to the non-unitary Lindblad operators. That is, any effect in the system sector that comes from integrating out the environment sector is visible only when we include the contribution from the environment sector in the interaction Hamiltonian.

\section{Interaction Hamiltonian for tensor perturbations}
\label{sec:Hint}
\setcounter{equation}{0}

Now we consider the interaction Hamiltonian for tensor perturbations. We consider as the leading higher-order interaction cubic order Hamiltonian that contains purely tensor perturbations $h_{ij}$. The cubic action for the tensor perturbation is~\cite{Prokopec:2010be,Gong:2016qpq}
\begin{align}
S_3^{(t)} 
& = 
\int d\tau d^3x a^2 \mpl^2 \bigg[
-\frac{1}{2} h_{ij}h_{jk}'h_{ki}' - 2\calH h_{ij}h_{jk}h_{ki}'
+ 2 \bigg( 1-\frac{\epsilon}{3} \bigg) \calH^2 h_{ij}h_{jk}h_{ki}
\nonumber\\
& \hspace{8em}
+ h_{ij} \bigg( \frac{1}{4} h_{kl,i}h_{kl,j} + \frac{1}{2}h_{ik,l} h_{jl,k}
- \frac{3}{2} h_{ik,l}h_{jk,l} \bigg) \bigg] 
\, ,
\end{align}
where $\calH \equiv a'/a$ and $\epsilon \equiv -\dot{H}/H^2$. There are two things to notice:
\begin{enumerate}

\item Some terms are not slow-roll suppressed, and
\item some terms do not possess any derivatives.

\end{enumerate}
Then, the interaction Hamiltonian we consider is
\begin{align}
\label{eq:Hint}
H_\text{int} & = -a^2\mpl^2 \int d^3x \bigg[
-\frac{1}{2} h_{ij}h_{jk}'h_{ki}' - 2\calH h_{ij}h_{jk}h_{ki}'
+ 2 \bigg( 1-\frac{\epsilon}{3} \bigg) \calH^2 h_{ij}h_{jk}h_{ki}
\nonumber\\
& \hspace{8em}
+ h_{ij} \bigg( \frac{1}{4} h_{kl,i}h_{kl,j} + \frac{1}{2}h_{ik,l} h_{jl,k}
- \frac{3}{2} h_{ik,l}h_{jk,l} \bigg) \bigg] \, .
\end{align}
As we are interested in the Fourier modes, going to the momentum space using the free solutions \eqref{eq:polarization-decompose}, \eqref{eq:canonicaltensor} and \eqref{eq:sol} gives the following explicit expression for the interaction Hamiltonian:
\begin{align}
\label{eq:Hint2}
H_\text{int} = &
\int \frac{d^3k_1}{(2\pi)^3} \frac{d^3k_2}{(2\pi)^3} \frac{d^3k_3}{(2\pi)^3}
(2\pi)^3 \delta^{(3)}({\bm k}_{123}) 
\sum_{\lambda_1,\lambda_2,\lambda_3} 
\nonumber\\
& \times \Big[ 
\big( c_1 + ic_{21} + ic_{22} + ic_{23} + c_{31} + c_{32} + c_{33} \big)
a_{{\bm k}_1}^{\lambda_1}a_{{\bm k}_2}^{\lambda_2}a_{{\bm k}_3}^{\lambda_3}
\nonumber\\
& \qquad
+ \big( c_1 - ic_{21} + ic_{22} + ic_{23} + c_{31} - c_{32} - c_{33} \big)
a_{-{\bm k}_1}^{\lambda_1\dag}a_{{\bm k}_2}^{\lambda_2}a_{{\bm k}_3}^{\lambda_3}
\nonumber\\
& \qquad
+ \big( c_1 + ic_{21} - ic_{22} + ic_{23} - c_{31} + c_{32} - c_{33} \big)
a_{{\bm k}_1}^{\lambda_1}a_{-{\bm k}_2}^{\lambda_2\dag}a_{{\bm k}_3}^{\lambda_3}
\nonumber\\
& \qquad
+ \big( c_1 + ic_{21} + ic_{22} - ic_{23} - c_{31} - c_{32} + c_{33} \big)
a_{{\bm k}_1}^{\lambda_1}a_{{\bm k}_2}^{\lambda_2}a_{-{\bm k}_3}^{\lambda_3\dag}
\nonumber\\
& \qquad
+ \big( c_1 + ic_{21} - ic_{22} - ic_{23} + c_{31} - c_{32} - c_{33} \big)
a_{{\bm k}_1}^{\lambda_1}a_{-{\bm k}_2}^{\lambda_2\dag}a_{-{\bm k}_3}^{\lambda_3\dag}
\nonumber\\
& \qquad
+ \big( c_1 - ic_{21} + ic_{22} - ic_{23} - c_{31} + c_{32} - c_{33} \big)
a_{-{\bm k}_1}^{\lambda_1\dag}a_{{\bm k}_2}^{\lambda_2}a_{-{\bm k}_3}^{\lambda_3\dag}
\nonumber\\
& \qquad
+ \big( c_1 - ic_{21} - ic_{22} + ic_{23} - c_{31} - c_{32} + c_{33} \big)
a_{-{\bm k}_1}^{\lambda_1\dag}a_{-{\bm k}_2}^{\lambda_2\dag}a_{{\bm k}_3}^{\lambda_3}
\nonumber\\
& \qquad
+ \big( c_1 - ic_{21} - ic_{22} - ic_{23} + c_{31} + c_{32} + c_{33} \big)
a_{-{\bm k}_1}^{\lambda_1\dag}a_{-{\bm k}_2}^{\lambda_2\dag}a_{-{\bm k}_3}^{\lambda_3\dag}
\Big] \, ,
\end{align}
where 
\begin{align}
\label{eq:coeff-c1}
c_1
& \equiv
\frac{1}{2a\mpl} \frac{1}{\sqrt{k_1k_2k_3}}
\bigg\{
- 4 \bigg( 1-\frac{\epsilon}{3} \bigg) \calH^2
e_{ij}^{\lambda_1}({\bm k}_1) e_{jk}^{\lambda_2}({\bm k}_2) e_{ki}^{\lambda_3}({\bm k}_3) 
\nonumber\\
& \quad
+ \frac{1}{6} \Big[
e_{ij}^{\lambda_1}({\bm k}_1) e_{kl}^{\lambda_2}({\bm k}_2) e_{kl}^{\lambda_3}({\bm k}_3) 
k_2^ik_3^j
+
e_{ij}^{\lambda_2}({\bm k}_2) e_{kl}^{\lambda_3}({\bm k}_3) e_{kl}^{\lambda_1}({\bm k}_1) 
k_3^ik_1^j
+
 e_{ij}^{\lambda_3}({\bm k}_3) e_{kl}^{\lambda_1}({\bm k}_1) e_{kl}^{\lambda_2}({\bm k}_2) 
k_1^ik_2^j \Big]
\nonumber\\
& \quad
+ \frac{1}{3} \Big[
e_{ij}^{\lambda_1}({\bm k}_1) e_{ik}^{\lambda_2}({\bm k}_2) e_{jl}^{\lambda_3}({\bm k}_3) 
k_2^lk_3^k
+ 
e_{ij}^{\lambda_2}({\bm k}_2) e_{ik}^{\lambda_3}({\bm k}_3) e_{jl}^{\lambda_1}({\bm k}_1) 
k_3^lk_1^k
+ 
e_{ij}^{\lambda_3}({\bm k}_3) e_{ik}^{\lambda_1}({\bm k}_1) e_{jl}^{\lambda_2}({\bm k}_2) 
k_1^lk_2^k \Big]
\nonumber\\
& \quad
- \Big[
e_{ij}^{\lambda_1}({\bm k}_1) e_{ik}^{\lambda_2}({\bm k}_2) e_{jk}^{\lambda_3}({\bm k}_3) 
{\bm k}_2\cdot{\bm k}_3
+
e_{ij}^{\lambda_2}({\bm k}_2) e_{ik}^{\lambda_3}({\bm k}_3) e_{jk}^{\lambda_1}({\bm k}_1) 
{\bm k}_3\cdot{\bm k}_1
+
e_{ij}^{\lambda_3}({\bm k}_3) e_{ik}^{\lambda_1}({\bm k}_1) e_{jk}^{\lambda_2}({\bm k}_2)
{\bm k}_1\cdot{\bm k}_2 \Big]
\bigg\}
\, ,
\\
\label{eq:coeff-c2}
c_{21}
& \equiv
-\frac{2}{3} \frac{\calH}{a\mpl} \sqrt{\frac{k_1}{k_2k_3}}
e_{ij}^{\lambda_1}({\bm k}_1) e_{jk}^{\lambda_2}({\bm k}_2) e_{ki}^{\lambda_3}({\bm k}_3) 
\text{ and similar for $c_{22}$ and $c_{23}$}
\, ,
\\
\label{eq:coeff-c3}
c_{31}
& \equiv
-\frac{1}{6a\mpl} \sqrt{\frac{k_2k_3}{k_1}}
e_{ij}^{\lambda_1}({\bm k}_1) e_{jk}^{\lambda_2}({\bm k}_2) e_{ki}^{\lambda_3}({\bm k}_3) 
\text{ and similar for $c_{32}$ and $c_{33}$}
\, .
\end{align}

Before we make further steps forward, let us investigate the time dependence of \eqref{eq:Hint2}. Here, the creation and annihilation operators $a_{{\bm k}_i}^{\lambda_i}(\tau)$ and $a_{-{\bm k}_i}^{\lambda_i\dag}(\tau)$ are coming from the mode function solution $h_\lambda({\bm k}_i,\tau)$, and are time-dependent -- this is clearly not the case for the Schr\"odinger picture. In fact, promoting the canonical conjugate pair $\pi_{\bm k}$ and $v_{\bm k}$ to time-dependent quantum operators as in \eqref{eq:variables} means we make use of the Heisenberg picture. The general transformation of an operator $A$ from the Schr\"odinger picture to the interaction picture is determined only by the free Hamiltonian as $A_I(t) = U_0^\dag(t;t_0) A_S U_0(t;t_0)$\footnote{
We may regard this as the evolution of an operator in the interaction picture, upon identifying the initial conditions such that
\begin{equation*}
A_I(\tau_0) = A_S \, .
\end{equation*}
If we are in the Heisenberg picture, the evolution of an operator -- or the transformation from the Schr\"odinger picture to the Heisenberg picture -- is determined by the {\it full} Hamiltonian, $A_H(\tau) = U^\dag(\tau;\tau_0) A_S U(\tau;\tau_0)$ with
\begin{equation*}
U(\tau;\tau_0) = T \exp \bigg[ -i \int_{\tau_0}^\tau H(\tau') d\tau' \bigg] \, .
\end{equation*}
}, 
also applies to the creation and annihilation operators as
\begin{equation}
\label{eq:operator-evolution}
a_{{\bm k},I}(\tau) = U_0^\dag(\tau;\tau_0) a_{\bm k}(\tau_0) U_0(\tau;\tau_0)
\end{equation}
and the same for $a_{-{\bm k}}^\dag(\tau)$, and the explicit form is given in terms of the Bogoliubov transformation \eqref{eq:bogoliubov}. That is, \eqref{eq:bogoliubov} tells us how the time-independent -- in the sense that it is defined at $\tau_0$ and is fixed there -- Schr\"odinger picture operators $a_{\bm k}(\tau_0)$ and $a_{-{\bm k}}^\dag(\tau_0)$ are related to the time-dependent interaction picture operators $a_{\bm k}(\tau) \equiv a_{{\bm k},I}(\tau)$ and $a_{-{\bm k}}^\dag(\tau) \equiv a_{-{\bm k},I}^\dag(\tau)$. Returning to \eqref{eq:Hint2}, as mentioned above the creation and annihilation operators are from $h_\lambda({\bm k}_i,\tau)$, where the relations to the operators at the initial moments are given by \eqref{eq:bogoliubov}. That is, the evolution of the creation and annihilation operators in \eqref{eq:Hint2} is determined by the quadratic Hamiltonian $H_0$, and thus \eqref{eq:Hint2} is in the interaction picture. Note that
\begin{align}
\label{eq:Hint4}
\hspace{-3em}
\eqref{eq:Hint2} & = H_{\text{int},I}(\tau)
\nonumber\\
& = 
\int  \Big\{ \big[ c_1(\tau) + \cdots \big] 
a_{{\bm k}_1,I}^{\lambda_1}(\tau) a_{{\bm k}_2,I}^{\lambda_2}(\tau) a_{{\bm k}_3,I}^{\lambda_3}(\tau) 
+ \cdots \Big\}
\nonumber\\
& = 
\int \bigg\{ \big[ c_1(\tau) + \cdots \big]  
\Big[ U_0^\dag(\tau;\tau_0) a_{{\bm k}_1}^{\lambda_1}(\tau_0) U_0(\tau;\tau_0) \Big]
\Big[ U_0^\dag(\tau;\tau_0) a_{{\bm k}_2}^{\lambda_2}(\tau_0) U_0(\tau;\tau_0) \Big]
\Big[ U_0^\dag(\tau;\tau_0) a_{{\bm k}_3}^{\lambda_3}(\tau_0) U_0(\tau;\tau_0) \Big] + \cdots 
\bigg\}
\nonumber\\
& =
U_0^\dag(\tau;\tau_0) \underbrace{ \int \Big\{ \big[ c_1(\tau) + \cdots \big] 
a_{{\bm k}_1}^{\lambda_1}(\tau_0) a_{{\bm k}_2}^{\lambda_2}(\tau_0) a_{{\bm k}_3}^{\lambda_3}(\tau_0) 
+ \cdots \Big\} }_{\equiv H_{\text{int},S}(\tau) } U_0(\tau;\tau_0) 
\, .
\end{align}
Here, $H_{\text{int},S}(\tau)$ is the interaction Hamiltonian in the Schr\"odinger picture in the sense that the operators are those at $\tau_0$, while the coefficients are evaluated at $\tau$ and the relation between them is given by the standard form: $H_{\text{int},I}(\tau) = U^\dag(\tau;\tau_0) H_{\text{int},S}(\tau) U(\tau;\tau_0)$. Another benefit to make use of the initial creation and annihilation operators at $\tau_0$ is that, after all, we wish to compute the time evolution of the reduced density matrix of the system $\rho_\text{red}$, for which we need to find the Lindblad operators \eqref{eq:Lindblad-op1} and \eqref{eq:Lindblad-op2}. As we can see, they are essentially the interaction Hamiltonian sandwiched between the time-evolved environment vacuum states. Thus, combined with time evolution operators accompanied by $H_{\text{int},I}$, the initial annihilation operators directly eliminate the initial environment vacuum state $|\calE(\tau_0)\rangle$, reducing amounts of computational efforts for the Lindblad operators.

While we have the formal relation between creation and annihilation operators at $\tau$ and those at $\tau_0$ as \eqref{eq:operator-evolution}, practically we can also use \eqref{eq:bogoliubov} conveniently to write 
\begin{align}
\label{eq:gamma-def}
a_{\bm k}(\tau) \pm a_{-{\bm k}}^{\dag}(\tau)
& = 
\Big[ \alpha_k(\tau) a_{\bm k}(\tau_0) + \beta_k(\tau) a_{-{\bm k}}^\dag(\tau) \Big]
\pm \Big[ \beta_k^*(\tau) a_{\bm k}(\tau_0) + \alpha_k^*(\tau) a_{-{\bm k}}^\dag(\tau_0) \Big]
\nonumber\\
& =
\Big[ \alpha_k(\tau) \pm \beta_k^*(\tau) \Big] a_{\bm k}(\tau_0)
\pm \Big[ \alpha_k^*(\tau) \pm \beta_k(\tau) \Big] a_{-{\bm k}}^\dag(\tau_0)
\nonumber\\
& \equiv
\gamma_{k\pm}(\tau) a_{\bm k}(\tau_0) \pm \gamma_{k\pm}^*(\tau) a_{-{\bm k}}^\dag(\tau_0) \, .
\end{align}
Then we can write \eqref{eq:Hint2} straightforwardly in terms of the combination of the coefficients at $\tau$ and the operators at $\tau_0$ as
\begin{align}
\label{eq:Hint5}
H_{\text{int},I}(\tau)
= &
\int \frac{d^3k_1}{(2\pi)^3} \frac{d^3k_2}{(2\pi)^3} \frac{d^3k_3}{(2\pi)^3}
(2\pi)^3 \delta^{(3)}({\bm k}_{123}) 
\sum_{\lambda_1,\lambda_2,\lambda_3} 
\nonumber\\
& \times 
\bigg[ \Big( c_1\gamma_{k_1+}^{\lambda_1}\gamma_{k_2+}^{\lambda_2}\gamma_{k_3+}^{\lambda_3} 
+ ic_{21}\gamma_{k_1-}^{\lambda_1}\gamma_{k_2+}^{\lambda_2}\gamma_{k_3+}^{\lambda_3} 
+ ic_{22}\gamma_{k_1+}^{\lambda_1}\gamma_{k_2-}^{\lambda_2}\gamma_{k_3+}^{\lambda_3} 
+ ic_{23}\gamma_{k_1+}^{\lambda_1}\gamma_{k_2+}^{\lambda_2}\gamma_{k_3-}^{\lambda_3}
\nonumber\\
& \qquad
+ c_{31}\gamma_{k_1+}^{\lambda_1}\gamma_{k_2-}^{\lambda_2}\gamma_{k_3-}^{\lambda_3} 
+ c_{32}\gamma_{k_1-}^{\lambda_1}\gamma_{k_2+}^{\lambda_2}\gamma_{k_3-}^{\lambda_3}
+ c_{33}\gamma_{k_1-}^{\lambda_1}\gamma_{k_2-}^{\lambda_2}\gamma_{k_3+}^{\lambda_3} \Big) 
a_{{\bm k}_1}^{\lambda_1}a_{{\bm k}_2}^{\lambda_2}a_{{\bm k}_3}^{\lambda_3}
\nonumber\\
& \quad
+ \Big( c_1\gamma_{k_1+}^{\lambda_1*}\gamma_{k_2+}^{\lambda_2}\gamma_{k_3+}^{\lambda_3} 
- ic_{21}\gamma_{k_1-}^{\lambda_1*}\gamma_{k_2+}^{\lambda_2}\gamma_{k_3+}^{\lambda_3}
+ ic_{22}\gamma_{k_1+}^{\lambda_1*}\gamma_{k_2-}^{\lambda_2}\gamma_{k_3+}^{\lambda_3} 
+ ic_{23}\gamma_{k_1+}^{\lambda_1*}\gamma_{k_2+}^{\lambda_2}\gamma_{k_3-}^{\lambda_3} 
\nonumber\\
& \qquad
+ c_{31}\gamma_{k_1+}^{\lambda_1*}\gamma_{k_2-}^{\lambda_2}\gamma_{k_3-}^{\lambda_3} 
- c_{32}\gamma_{k_1-}^{\lambda_1*}\gamma_{k_2+}^{\lambda_2}\gamma_{k_3-}^{\lambda_3}
- c_{33}\gamma_{k_1-}^{\lambda_1*}\gamma_{k_2-}^{\lambda_2}\gamma_{k_3+}^{\lambda_3} \Big) 
a_{-{\bm k}_1}^{\lambda_1\dag} a_{{\bm k}_2}^{\lambda_2} a_{{\bm k}_3}^{\lambda_3}
\nonumber\\
& \quad
+ \Big( c_1\gamma_{k_1+}^{\lambda_1}\gamma_{k_2+}^{\lambda_2*}\gamma_{k_3+}^{\lambda_3} 
+ ic_{21}\gamma_{k_1-}^{\lambda_1}\gamma_{k_+}^{\lambda_2*}\gamma_{k_3+}^{\lambda_3}
- ic_{22}\gamma_{k_1+}^{\lambda_1}\gamma_{k_-}^{\lambda_2*}\gamma_{k_3+}^{\lambda_3} 
+ ic_{23}\gamma_{k_1+}^{\lambda_1}\gamma_{k_2+}^{\lambda_2*}\gamma_{k_3-}^{\lambda_3}
\nonumber\\
& \qquad
- c_{31}\gamma_{k_1+}^{\lambda_1}\gamma_{k_2-}^{\lambda_2*}\gamma_{k_3-}^{\lambda_3} 
+ c_{32}\gamma_{k_1-}^{\lambda_1}\gamma_{k_2+}^{\lambda_2*}\gamma_{k_3-}^{\lambda_3}
- c_{33}\gamma_{k_1-}^{\lambda_1}\gamma_{k_2-}^{\lambda_2*}\gamma_{k_3+}^{\lambda_3} \Big) 
a_{{\bm k}_1}^{\lambda_1}a_{-{\bm k}_2}^{\lambda_2\dag} a_{{\bm k}_3}^{\lambda_3}
\nonumber\\
& \quad
+ \Big( c_1\gamma_{k_1+}^{\lambda_1}\gamma_{k_2+}^{\lambda_2}\gamma_{k_3+}^{\lambda_3*} 
+ ic_{21}\gamma_{k_1-}^{\lambda_1}\gamma_{k_2+}^{\lambda_2}\gamma_{k_3+}^{\lambda_3*}
+ ic_{22}\gamma_{k_1+}^{\lambda_1}\gamma_{k_2-}^{\lambda_2}\gamma_{k_3+}^{\lambda_3*} 
- ic_{23}\gamma_{k_1+}^{\lambda_1}\gamma_{k_2+}^{\lambda_1}\gamma_{k_3-}^{\lambda_3*}
\nonumber\\
& \qquad
- c_{31}\gamma_{k_1+}^{\lambda_1}\gamma_{k_2-}^{\lambda_2}\gamma_{k_3-}^{\lambda_3*} 
- c_{32}\gamma_{k_1-}^{\lambda_1}\gamma_{k_2+}^{\lambda_2}\gamma_{k_3-}^{\lambda_3*}
+ c_{33}\gamma_{k_1-}^{\lambda_1}\gamma_{k_2-}^{\lambda_2}\gamma_{k_3+}^{\lambda_3*} \Big) 
a_{{\bm k}_1}^{\lambda_1}a_{{\bm k}_2}^{\lambda_2}a_{-{\bm k}_3}^{\lambda_3\dag}
\nonumber\\
& \quad
+ \Big( c_1\gamma_{k_1+}^{\lambda_1}\gamma_{k_2+}^{\lambda_2*}\gamma_{k_3+}^{\lambda_3*} 
+ ic_{21}\gamma_{k_1-}^{\lambda_1}\gamma_{k_2+}^{\lambda_2*}\gamma_{k_3+}^{\lambda_3*}
- ic_{22}\gamma_{k_1+}^{\lambda_1}\gamma_{k_2-}^{\lambda_2*}\gamma_{k_3+}^{\lambda_3*} 
- ic_{23}\gamma_{k_1+}^{\lambda_1}\gamma_{k_2+}^{\lambda_2*}\gamma_{k_3-}^{\lambda_3*} 
\nonumber\\
& \qquad
+ c_{31}\gamma_{k_1+}^{\lambda_1}\gamma_{k_2-}^{\lambda_2*}\gamma_{k_3-}^{\lambda_3*} 
- c_{32}\gamma_{k_1-}^{\lambda_1}\gamma_{k_2+}^{\lambda_2*}\gamma_{k_3-}^{\lambda_3*}
- c_{33}\gamma_{k_1-}^{\lambda_1}\gamma_{k_2-}^{\lambda_2*}\gamma_{k_3+}^{\lambda_3*} \Big) 
a_{{\bm k}_1}^{\lambda_1}a_{-{\bm k}_2}^{\lambda_2\dag}a_{-{\bm k}_3}^{\lambda_3\dag}
\nonumber\\
& \quad
+ \Big( c_1\gamma_{k_1+}^{\lambda_1*}\gamma_{k_2+}^{\lambda_2}\gamma_{k_3+}^{\lambda_3*} 
- ic_{21}\gamma_{k_1-}^{\lambda_1*}\gamma_{k_2+}^{\lambda_2}\gamma_{k_3+}^{\lambda_3*}
+ ic_{22}\gamma_{k_1+}^{\lambda_1*}\gamma_{k_2-}^{\lambda_2}\gamma_{k_3+}^{\lambda_3*} 
- ic_{23}\gamma_{k_1+}^{\lambda_1*}\gamma_{k_2+}^{\lambda_2}\gamma_{k_3-}^{\lambda_3*}
\nonumber\\
& \qquad
- c_{31}\gamma_{k_1+}^{\lambda_1*}\gamma_{k_2-}^{\lambda_2}\gamma_{k_3-}^{\lambda_3*} 
+ c_{32}\gamma_{k_1-}^{\lambda_1*}\gamma_{k_2+}^{\lambda_2}\gamma_{k_3-}^{\lambda_3*}
- c_{33}\gamma_{k_1-}^{\lambda_1*}\gamma_{k_2-}^{\lambda_2}\gamma_{k_3+}^{\lambda_3*} \Big) 
a_{-{\bm k}_1}^{\lambda_1\dag}a_{{\bm k}_2}^{\lambda_2}a_{-{\bm k}_3}^{\lambda_3\dag}
\nonumber\\
& \quad
+ \Big( c_1\gamma_{k_1+}^{\lambda_1*}\gamma_{k_2+}^{\lambda_2*}\gamma_{k_3+}^{\lambda_3} 
- ic_{21}\gamma_{k_1-}^{\lambda_1*}\gamma_{k_2+}^{\lambda_2*}\gamma_{k_3+}^{\lambda_3}
- ic_{22}\gamma_{k_1+}^{\lambda_1*}\gamma_{k_2-}^{\lambda_2*}\gamma_{k_3+}^{\lambda_3} 
+ ic_{23}\gamma_{k_1+}^{\lambda_1*}\gamma_{k_2+}^{\lambda_2*}\gamma_{k_3-}^{\lambda_3}
\nonumber\\
& \qquad
- c_{31}\gamma_{k_1+}^{\lambda_1*}\gamma_{k_2-}^{\lambda_2*}\gamma_{k_3-}^{\lambda_3} 
- c_{32}\gamma_{k_1-}^{\lambda_1*}\gamma_{k_2+}^{\lambda_2*}\gamma_{k_3-}^{\lambda_3}
+ c_{33}\gamma_{k_1-}^{\lambda_1*}\gamma_{k_2-}^{\lambda_2*}\gamma_{k_3+}^{\lambda_3} \Big) 
a_{{\bm k}_1}^{\lambda_1\dag}a_{-{\bm k}_2}^{\lambda_2\dag}a_{{\bm k}_3}^{\lambda_3}
\nonumber\\
& \quad
+ \Big( c_1\gamma_{k_1+}^{\lambda_1*}\gamma_{k_2+}^{\lambda_2*}\gamma_{k_3+}^{\lambda_3*} 
- ic_{21}\gamma_{k_1-}^{\lambda_1*}\gamma_{k_2+}^{\lambda_2*}\gamma_{k_3+}^{\lambda_3*}
- ic_{22}\gamma_{k_1+}^{\lambda_1*}\gamma_{k_2-}^{\lambda_2*}\gamma_{k_3+}^{\lambda_3*} 
- ic_{23}\gamma_{k_1+}^{\lambda_1*}\gamma_{k_2+}^{\lambda_2*}\gamma_{k_3-}^{\lambda_3*}
\nonumber\\
& \qquad
+ c_{31}\gamma_{k_1+}^{\lambda_1*}\gamma_{k_2-}^{\lambda_2*}\gamma_{k_3-}^{\lambda_3*} 
+ c_{32}\gamma_{k_1-}^{\lambda_1*}\gamma_{k_2+}^{\lambda_2*}\gamma_{k_3-}^{\lambda_3*}
+ c_{33}\gamma_{k_1-}^{\lambda_1*}\gamma_{k_2-}^{\lambda_2*}\gamma_{k_3+}^{\lambda_3*} \Big) 
a_{{\bm k}_1}^{\lambda_1\dag}a_{-{\bm k}_2}^{\lambda_2\dag}a_{-{\bm k}_3}^{\lambda_3\dag}
\bigg] 
\, .
\end{align}

One more simplification is to come. The real condition of the fluctuations demands the following reality condition on the polarization tensor:
\begin{equation}
e_{ij}^{\lambda*}({\bm k}) = e_{ij}^\lambda(-{\bm k}) \, .
\end{equation}
Since the three-momenta are being integrated, the opposite sign of the momentum combined with an annihilation operator gives
\begin{equation}
\label{eq:k-hc}
\int \frac{d^3k}{(2\pi)^3} \Big[ e_{ij}^{\lambda}({\bm k}) a_{\bm k}^\lambda \Big]^\dag
= \int \frac{d^3k}{(2\pi)^3} e_{ij}^\lambda(-{\bm k}) a_{{\bm k}}^{\lambda\dag}
= \int \frac{d^3q}{(2\pi)^3} e_{ij}^\lambda({\bm q}) a_{-{\bm q}}^{\lambda\dag} \, ,
\end{equation}
where for the last equality we have set ${\bm q} \equiv -{\bm k}$. Since the coefficients $c_{i(j)}$ in \eqref{eq:Hint2} are basically combinations of polarization tensor, we can see for example
\begin{align}
\int \Big[ -ic_{21}({\bm k}_1,{\bm k}_2,{\bm k}_3) a_{-{\bm k}_1}^{\lambda_1\dag} 
a_{{\bm k}_2}^{\lambda_2} a_{{\bm k}_3}^{\lambda_2} \Big]^\dag
& = \int i \frac{-2}{3}\frac{\calH}{a\mpl} e_{ij}({\bm k}_2)e_{jk}({\bm k}_3)e_{ki}({\bm k}_1) \sqrt{\frac{k_1}{k_2k_3}}
a^{\lambda_3\dag}_{-{\bm k}_3}a^{\lambda_2\dag}_{-{\bm k}_2}a^{\lambda_1}_{{\bm k}_1}
\nonumber\\
& = 
\int i \frac{-2}{3}\frac{\calH}{a\mpl} e_{ij}({\bm k}_2)e_{jk}({\bm k}_1')e_{ki}({\bm k}_3') \sqrt{\frac{k_3'}{k_2k_1'}}
a^{\lambda_{1'}\dag}_{-{\bm k}_1'}a^{\lambda_2\dag}_{-{\bm k}_2}a^{\lambda_{3'}}_{{\bm k}_3'}
\nonumber\\
& = 
\int ic_{23} a^{\lambda_1\dag}_{-{\bm k}_1}a^{\lambda_2\dag}_{-{\bm k}_2}a^{\lambda_3}_{{\bm k}_3} 
\, ,
\end{align}
where for the last equality we have renamed the dummy integration momentum variables as ${\bm k}_3 \equiv {\bm k}_1'$ and ${\bm k}_1 = {\bm k}_3'$. Similar relations holds for other coefficients and other operator combinations, e.g. $\int \big[ c_{31} a_{{\bm k}_1}a_{-{\bm k}_2}^\dag a_{{\bm k}_3} \big]^\dag = \int c_{33} a^\dag_{-{\bm k}_1}a_{{\bm k}_2}a^\dag_{-{\bm k}_3}$. Then we can see that the Hermitian conjugate pair of each term of \eqref{eq:Hint2} can be written as one another term, e.g.
\begin{align}
& \int \Big[ \big( c_1 - ic_{21} + ic_{22} + ic_{23} + c_{31} - c_{32} - c_{33} \big)
a_{-{\bm k}_1}^{\lambda_1\dag}a_{{\bm k}_2}^{\lambda_2}a_{{\bm k}_3}^{\lambda_3} \Big]^\dag
\nonumber\\
& =
\int \Big( c_1 + ic_{23} - ic_{22} - ic_{21} + c_{33} - c_{32} - c_{31} \big)
a_{-{\bm k}_1}^{\lambda_1\dag}a_{-{\bm k}_2}^{\lambda_2\dag}a_{{\bm k}_3}^{\lambda_3} \, .
\end{align}
Furthermore, since the structure of $c_{2i}$ and $c_{3i}$ is of the form
\begin{equation}
c_{2i} \sim \sqrt{k_i} \quad \text{and} \quad c_{3i} \sim \frac{1}{\sqrt{k_i}} \, ,
\end{equation}
shuffling the dummy momentum variables shuffles these coefficients, e.g. ${\bm k}_2 \to {\bm k}_1$, ${\bm k}_3 \to {\bm k}_2$ and ${\bm k}_1 \to {\bm k}_3$ gives
\begin{equation}
\begin{split}
& c_{21} \to c_{23} \, , \quad c_{22} \to c_{21} \quad \text{and} \quad c_{23} \to c_{22} \, ,
\\
& c_{31} \to c_{33} \, , \quad c_{32} \to c_{31} \quad \text{and} \quad c_{33} \to c_{32} \, ,
\end{split}
\end{equation}
and correspondingly under these changes
\begin{align}
&
\int \big( c_1 + ic_{21} - ic_{22} + ic_{23} - c_{31} + c_{32} - c_{33} \big)
a_{{\bm k}_1}^{\lambda_1}a_{-{\bm k}_2}^{\lambda_2\dag}a_{{\bm k}_3}^{\lambda_3}
\nonumber\\
& =
\int \big( c_1 + ic_{23} - ic_{21} + ic_{22} - c_{33} + c_{31} - c_{32} \big)
a_{{\bm k}_3}^{\lambda_3}a_{-{\bm k}_1}^{\lambda_1\dag}a_{{\bm k}_2}^{\lambda_2} \, ,
\end{align}
where now the coefficient is identical to that of $a_{-{\bm k}_1}^{\lambda_1\dag}a_{{\bm k}_2}^{\lambda_2}a_{{\bm k}_3}^{\lambda_3}$ in \eqref{eq:Hint2}. The same holds for $a_{{\bm k}_1}^{\lambda_1}a_{{\bm k}_2}^{\lambda_2}a_{-{\bm k}_3}^{\lambda_3\dag}$ under ${\bm k}_3 \to {\bm k}_1$, ${\bm k}_1 \to {\bm k}_2$ and ${\bm k}_2 \to {\bm k}_3$ in \eqref{eq:Hint2}. Thus, finally we can write the interaction Hamiltonian $H_\text{int}$ very simply as
\begin{align}
\label{eq:Hint3}
H_{\text{int},I}(\tau)
& = 
\int \frac{d^3k_1}{(2\pi)^3} \frac{d^3k_2}{(2\pi)^3} \frac{d^3k_3}{(2\pi)^3}
(2\pi)^3 \delta^{(3)}({\bm k}_{123}) 
\sum_{\lambda_1,\lambda_2,\lambda_3} 
\nonumber\\
& 
\quad
\times
\bigg\{ h_0(\tau) a_{{\bm k}_1}^{\lambda_1}a_{{\bm k}_2}^{\lambda_2}a_{{\bm k}_3}^{\lambda_3}(\tau_0)
+ h_1(\tau) \Big[
a_{-{\bm k}_1}^{\lambda_1\dag} a_{{\bm k}_2}^{\lambda_2} a_{{\bm k}_3}^{\lambda_3}(\tau_0)
+ a_{{\bm k}_3}^{\lambda_3}a_{-{\bm k}_1}^{\lambda_1\dag} a_{{\bm k}_2}^{\lambda_2}(\tau_0)
+ a_{{\bm k}_2}^{\lambda_2}a_{{\bm k}_3}^{\lambda_3} a_{-{\bm k}_1}^{\lambda_1\dag}(\tau_0)
\Big]
+ h.c.
\bigg\}
\, ,
\end{align}
where $h_0$ and $h_1$ are the coefficients of $a_{{\bm k}_1}^{\lambda_1}a_{{\bm k}_2}^{\lambda_2}a_{{\bm k}_3}^{\lambda_3}$ and $a_{-{\bm k}_1}^{\lambda_1\dag} a_{{\bm k}_2}^{\lambda_2} a_{{\bm k}_3}^{\lambda_3}$ in \eqref{eq:Hint5} respectively. Note that in the last step we have written the half of $H_{\text{int},I}$ as the Hermitian conjugates of the other half.
As emphasized before and written explicitly above, the coefficients $h_0$ and $h_1$ are at $\tau$ yet the operators are at $\tau_0$, viz. $h_0 = h_0(\tau)$ and $a_{{\bm k}_1}^{\lambda_1}a_{{\bm k}_2}^{\lambda_2}a_{{\bm k}_3}^{\lambda_3} = a_{{\bm k}_1}^{\lambda_1}a_{{\bm k}_2}^{\lambda_2}a_{{\bm k}_3}^{\lambda_3}(\tau_0)$ and so on.

\section{Structure of Lindblad operators}
\label{sec:structure}
\setcounter{equation}{0}

\subsection{Contributions of the environment and system sectors}

Having found the explicit form of the interaction Hamiltonian \eqref{eq:Hint3}, now we can compute the Lindblad operators explicitly. Before then, however, it is helpful to overview their structure rather than getting involved detailed calculations. For this purpose, we need to go back to the point how the Lindblad operators are constructed. As we can see from \eqref{eq:Lindblad-op1} and \eqref{eq:Lindblad-op2}, the Lindblad operators $L_1$ and $L_2$ are the interaction Hamiltonian sandwiched between the time-evolved initial environment states $|\calE(\tau_0)\rangle$ and the final environment states $|\calE_i\rangle$ which we want to trace out. For example, we may choose $|\calE(\tau_0)\rangle$ in such a way that $|\calE(\tau_0)\rangle$ denotes the sub-horizon modes at the moment $\tau$. Then, as \eqref{eq:BDvacuum}, starting from the standard Bunch-Davies vacuum $|0\rangle_0$ where there is no excited particle for all ${\bm k}$, we may split $|0\rangle_0$ as a product of the infrared and ultraviolet parts: 
\begin{equation}
|0\rangle_0 
= 
|0\rangle_\text{$k<aH$ at $\tau$} 
\otimes |0\rangle_\text{$k>aH$ at $\tau$} 
\equiv |0\rangle_\calS \otimes |0\rangle_\calE \, .
\end{equation}
That is, instead of writing the initial states for the system and environment respectively as $|\calS(\tau_0)\rangle$ and $|\calE(\tau_0)\rangle$, we now write them as $|0\rangle_\calS$ and $|0\rangle_\calE$ since we take the standard assumption that the initial vacuum is the Bunch-Davies one $|0\rangle_0$ for all modes. The infrared and ultraviolet vacua $|0\rangle_\calS$ and $|0\rangle_\calE$ are annihilated respectively by the annihilation operators of the infrared and ultraviolet modes, $a_{{\bm k}_\calS}$ and $a_{{\bm k}_\calE}$ where 
\begin{equation}
k_\calS \leq k = aH \text{ and } k_\calE \geq k = aH \text{ at } \tau \, .
\end{equation}
Thus separating the infrared and ultraviolet sectors of the interaction Hamiltonian enables us to treat one independent from the other. Consider, for example, $H_{\text{int},I}(\tau)$ is of the following form:
\begin{equation}
H_{\text{int},I}(\tau) =\int \frac{d^3k}{(2\pi)^3} h(\tau) a_{{\bm k}}(\tau_0) \, .
\end{equation}
Upon applying the ultraviolet vacuum state $|0\rangle_\calE$, the corresponding ultraviolet annihilation operators are operational and there are only left the annihilation operators for the infrared modes:
\begin{equation}
\label{eq:operator_UV-IR}
\int \frac{d^3k}{(2\pi)^3} a_{{\bm k}} |0\rangle_\calE 
= \bigg[ \int_{{\bm k} \in {\bm k}_\calS} \frac{d^3k}{(2\pi)^3} a_{{\bm k} \in {\bm k}_\calS} +
\int_{{\bm k} \in {\bm k}_\calE} \frac{d^3k}{(2\pi)^3} a_{{\bm k} \in {\bm k}_\calE} \bigg] |0\rangle_\calE
= \int_{{\bm k} \in {\bm k}_\calS} \frac{d^3k}{(2\pi)^3} a_{{\bm k} \in {\bm k}_\calS} |0\rangle_\calE \, .
\end{equation}
Further, the remaining annihilation operators for the infrared modes are linearly independent from the ultraviolet ones, so they can change their position regarding the ultraviolet annihilation operators. Thus
\begin{equation}
H_{\text{int},I}(\tau) |0\rangle_\calE = |0\rangle_\calE h(\tau) a_{{\bm k} \in {\bm k}_\calS} \, .
\end{equation}
That is, the infrared sector of $H_{\text{int},I}(\tau)$ is free from the ultraviolet vacuum $|0\rangle_\calE$.

More generally, based on the form \eqref{eq:Hint5}, let us schematically write $H_{\text{int},I}(\tau)$ as
\begin{equation}
\label{eq:Hint_UV-IR}
H_{\text{int},I}(\tau) = h^{mn}(\tau) H_{\calE,m}(\tau_0) H_{\calS,n}(\tau_0) \, ,
\end{equation}
where the overall coefficient is at $\tau$, and the system (or infrared) and environment (or ultraviolet) sectors of $H_{\text{int},I}$, which are combinations of the three initial creation and annihilation operators, are at $\tau_0$ so only the environment sector $H_{\calE,m}(\tau_0)$ is responding to the initial environment vacuum $|0\rangle_\calE$. It is very important to note that the right-hand side of \eqref{eq:Hint_UV-IR} denotes a single representative term in $H_{\text{int},I}$, and is {\it not} Hermitian, as clear from \eqref{eq:Hint5} -- while $H_{\text{int},I}$ itself is Hermitian also as clear from the same equation. In fact, to maintain the Hermiticity of $H_{\text{int},I}$ we had better always think of not a single term but a single Hermitian pair. But for brevity we keep writing a single term as \eqref{eq:Hint_UV-IR} while keeping in mind that we should remember the right-hand side of \eqref{eq:Hint_UV-IR} is not Hermitian. Then
\begin{equation}
L_\mu^\dag L_\nu 
\sim 
\langle0|_\calE H_{\text{int},I}^\dag(\tau_1) H_{\text{int},I}(\tau_2) |0\rangle_\calE 
= h^{mn*}(\tau_1) h^{pq}(\tau_2) 
\big\langle0\big|_\calE H_{\calE,m}^\dag(\tau_0)H_{\calE,p}(\tau_0) \big|0\big\rangle_\calE 
H_{\calS,n}^\dag(\tau_0) H_{\calS,q}(\tau_0) \, .
\end{equation}
While $H_{\text{int},I}$ is Hermitian, to keep the Hermitian-conjugateness of the Lindblad operators we write the Hermitian conjugate symbol for $H_{\text{int},I}$ as $H_{\text{int},I}^\dag$. The part sandwiched between the environment vacuum states above is some number -- exactly speaking, some function of momenta, including zero as well. Note that for the Lindblad operator terms in \eqref{eq:Lindblad}, as \eqref{eq:Hint_UV-IR}, we may decompose
\begin{equation}
\label{eq:Hint_UV-IR2}
\int_{\tau_0}^\tau d\tau_1 H_{\text{int},I}(\tau_1)
= g^{mn}(\tau) G_{\calE,m}(\tau_0) G_{\calS,n}(\tau_0) \, ,
\end{equation}
since the time integration only acts on the coefficients: $g^{mn}(\tau) \equiv \int_{\tau_0}^\tau d\tau_1 h^{mn}(\tau_1)$. Here, we have intentionally used different notations for the environment (system) sector as $G_{\calE(\calS),m}$, despite of the same time dependence at $\tau_0$, to avoid confusion in ordering the operators in $L_1$ (corresponding to $H$) and $L_2$ (corresponding to $G$). After simple calculations, we can schematically write the Lindblad terms of \eqref{eq:Lindblad} as
\begin{align}
\label{eq:Lindblad-terms}
& 
- \frac{1}{2} \sum_i \Big[
L_1^\dag L_2 \rho_{\calS}(\tau) + \rho_{\calS}(\tau)L_2^\dag L_1
- 2L_1\rho_{\calS}(\tau)L_2^\dag + (L_1 \leftrightarrow L_2) 
\Big]
\nonumber\\
= & 
-\frac{1}{2} \bigg\{
h^{mn*}(\tau)g^{pq}(\tau) 
\Big\langle0\Big|_\calE H_{\calE,m}^\dag(\tau_0) G_{\calE,p}(\tau_0) \Big|0\Big\rangle_\calE
\nonumber\\
& 
\quad\quad 
\times \Big[
U_{0,\calS}(\tau;\tau_0) H_{\calS,n}^\dag(\tau_0) G_{\calS,q}(\tau_0)
U_{0,\calS}^\dag(\tau;\tau_0) \rho_\calS(\tau)
+ \rho_\calS(\tau) U_{0,\calS}(\tau;\tau_0) H_{\calS,n}^\dag(\tau_0) G_{\calS,q}(\tau_0)
U_{0,\calS}^\dag(\tau;\tau_0)
\nonumber\\
& 
\qquad\quad
- 2 U_{0,\calS}(\tau;\tau_0) G_{\calS,q}(\tau_0) U_{0,\calS}^\dag(\tau;\tau_0) \rho_\calS(\tau)
U_{0,\calS}(\tau;\tau_0) H_{\calS,n}^\dag(\tau_0) U_{0,\calS}^\dag(\tau;\tau_0)
\Big]
+ h.c. \bigg\} 
\, .
\end{align}

To compute the environment parts of the Lindblad terms, naively thinking, it seems that only the operator combinations that consists of three annihilation operators first and three creation operators next survive, e.g.
\begin{equation}
\Big\langle0\Big|_\calE \Big( a_1a_2a_3 \Big)_\calE 
\Big( a_4^\dag a_5^\dag a_6^\dag \Big)_\calE \Big|0\Big\rangle_\calE 
\, .
\end{equation}
But given that the environment and system sectors of an operator are decomposed as \eqref{eq:operator_UV-IR}, it needs more care. There are surviving contributions in every term of \eqref{eq:Hint3}. For example, the first term of \eqref{eq:Hint3} contains no creation operator, so it seems that if both operator sets are of this type there would be no remaining contribution. But we do have a surviving contribution: with $\calO_\calS$ being the system operators not sandwiched between $|0\rangle_\calE$ like $\rho_\calS$ and $U_{0,\calS}(\tau;\tau_0)$, then even if both two operator sets consist of annihilation operators only, we have
\begin{align}
\label{eq:only-annihilation}
& \int \big\langle0\big|_\calE \big( a_1a_2a_3 \big) \big( a_4a_5a_6 \big) \big|0\big\rangle_\calE \calO_\calS
\nonumber\\
& = 
\int \big\langle0\big|_\calE \big[ (a_{1,\calS}+a_{1,\calE}) (a_{2,\calS}+a_{2,\calE}) \cdots \big] \big|0\big\rangle_\calE \calO_\calS
\nonumber\\
& =
\int \big\langle0\big|_\calE \big( a_{1,\calE}a_{2,\calE}a_{3,\calE} \big) 
\big( a_{4,\calE}a_{5,\calE}a_{6,\calE} \big) \big|0\big\rangle_\calE \calO_\calS
+ \cdots
+ \int \big\langle0\big|_\calE \big( a_{1,\calS}a_{2,\calS}a_{3,\calS} \big) 
\big( a_{4,\calS}a_{5,\calS}a_{6,\calS} \big) \big|0\big\rangle_\calE \calO_\calS
\nonumber\\
& =
\int_\calS \big( a_{1,\calS}a_{2,\calS}a_{3,\calS} \big) 
\big( a_{4,\calS}a_{5,\calS}a_{6,\calS} \big) \calO_\calS \, ,
\end{align}
where now the momentum integrations are only restricted to the system domain. But as argued in \eqref{eq:Hint-onlysystem}, those contributions in which the interaction Hamiltonian belongs only to the system sector can be absorbed into the unitary evolution and do not contribute to the Lindblad operators, so we do not consider such terms.

After taking into account all possible contractions between operators in the environment sector, the contributions appear as momentum-dependent (and also polarization-dependent) coefficients in front of the system operators. Thus the naive density matrix elements of system do change under the influence of the environment sector, as is what the Lindblad operators describe. But considering the environment sectors only does not seem to tell us any more, e.g. the structure of the time-dependent coefficients like $h_0(\tau)$, since what is constrained from the environment sectors are:
\begin{itemize}

\item Some momenta in the interaction Hamiltonian in $L_1$ and $L_2$ are related, e.g. ${\bm k}_1 = -{\bm k}_4$, and

\item certain momenta belong to certain domains, e.g. ${\bm k}_1$ and ${\bm k}_2$ are environmental (i.e. $|{\bm k}_1| \gg aH$ and $|{\bm k}_2| \gg aH$ at $\tau$) while ${\bm k}_3$ belong to the system sector (i.e. $|{\bm k}_3| \ll aH$ at $\tau$) and so on.

\end{itemize}
Thus, to make steps forward, we have to consider explicitly the system sector contributions. As we have started for the environment sector with the Bunch-Davies vacuum $|0\rangle_\calE$, so is the system vacuum $|0\rangle_\calS$. With $\rho_\calS(\tau)$ given by \eqref{eq:rho-freeS}, as we can see in \eqref{eq:Lindblad-terms} the evolution operator $U_{0,\calS}^\dag$ ($U_{0,\calS}$) in front of (behind) $\rho_\calS{(\tau)}$ cancels the counterpart in $\rho_\calS{(\tau)}$, so the initial vacuum state $|0\rangle_\calS$ ($\langle0|_\calS$) is exposed. Then the initial system creation or annihilation operators at $\tau_0$ from the Lindblad operators can act directly on $|0\rangle_\calS$.

During the computation in the system sector, it is convenient to arrange the Lindblad terms in their simplest form in the sense that there is no operator or are only creation ones on the left of $|0\rangle_\calS$, and likewise there is no operator or are only annihilation operators on the right of $\langle0|_\calS$, i.e. each term in ${d\rho_\text{red}}/{d\tau}$ is of schematic form
\begin{equation}
\label{eq:schematic-Lindblad-matrix}
\frac{d\rho_\text{red}}{d\tau} 
=
\rho_{mn} 
U_{0,\calS} a^\dag_{1,\calS}a^\dag_{2,\calS}\cdots a^\dag_{m,\calS} |0\rangle_\calS
\langle0|_\calS a_{1',\calS}a_{2',\calS}\cdots a_{n,\calS} U_{0,\calS}^\dag 
\, ,
\end{equation}
where $\rho_{mn}$ is the time and momentum dependent coefficient for the contribution with $m$ creation and $n$ annihilation operators, with $m, n \leq 6$. This form of the Lindblad term is irreducible, and applying the basis vectors on the left and right should give us the matrix elements in such bases directly.

\subsection{Matrix elements of reduced density matrix}

Now, we need to specify the ``matrix bases'' and to compute the matrix elements of the reduced density matrix with respect to those bases, or the states relevant for observations. The real issue here is, rather than computing the matrix elements which is quite straightforward, identifying what should be the observationally relevant states. Usually, the coherent state as an eigenstate of the annihilation operator is considered to be relevant for two reasons. First, it is the state with which the expectation values of operators follow the classical equation of motion. Second, the uncertainty is minimized in the coherent state. Thus, the coherent state coincides with our intuition of the classical solution. Especially, by taking expectation value of the field operators in terms of the coherent state\footnote{
Obviously, the coherent state is different from an eigenstate of the field operator. The reason we can take $\langle\pi_\phi\rangle = 0$ is: we are dealing with real fields so the expectation value of the operator $\pi_\phi$ vanishes, since $\pi_\phi \sim a - a^\dag$ [see \eqref{eq:sol}] and if $a$ has a real eigenvalue, $a^\dag$ in the expectation value is given by the same eigenvalue. Furthermore, uncertainty is still non-zero even if it is minimized, which means that as soon as we consider the eigenstate of the field operator, canonical momentum probability distribution spreads so we can no longer fix $\pi_\phi = 0$.
}, we can treat any operator in the Hamiltonian, say $\widehat{\calA}_i$, as a function of classical fields:
\begin{equation}
\widehat{\calA}_i \big( \hat{\phi},\hat{\pi}_\phi \big) 
\approx 
\calA_i(\phi,\pi_\phi) 
\, ,
\end{equation}
where we distinguish the field operators (hatted) and the classical fields (unhatted).

But as we have seen, the Lindblad terms are after computing out the system sector written in terms of the squeezed states: 
\begin{equation}
U_{0,\calS}|0\rangle_\calS \sim \sum_{n=0}^\infty |n,{\bm k}_\calS;n,-{\bm k}_\calS;\lambda\rangle
\quad \text{where} \quad
|n,{\bm k}_\calS;n,-{\bm k}_\calS;\lambda\rangle 
= \frac{1}{n!} \Big( a_{-{\bm k}_\calS}^{\lambda\dag} a_{{\bm k}_\calS}^{\lambda\dag} \Big)^n
|0\rangle_\calS \, .
\end{equation}
That is, the Lindblad terms are schematically of the form \eqref{eq:schematic-Lindblad-matrix}. Thus the squeezed states are already written in terms of the basis
\begin{equation}
\label{eq:matrixbases}
\Big\{ U_{0,\calS}|0\rangle_\calS, U_{0,\calS}a^\dag_{1,\calS}|0\rangle, 
U_{0,\calS}a^\dag_{1,\calS}a^\dag_{2,\calS}|0\rangle_\calS, \cdots \Big\} \, ,
\end{equation}
and in this sense the squeezed states are the natural bases so that the calculations become particularly easy if we adopt the squeezed states as the pointer bases\footnote{
Unlike the basis for the environment discussed in Section~\ref{sec:reduced}, the squeezed state basis is time dependent. However, such a time dependence corresponds to the unitary evolution of the Fock space states, so it is irrelevant to the non-unitary evolution through the Lindblad terms.
}. But obviously, this never means that we should adopt the squeezed states as the pointer bases. Rather, it just means that by adopting the squeezed states calculations becomes especially trivial.

One possible argument in favour of the squeezed states as the pointer bases is from noticing that we do not observe directly primordial perturbations. What we do observe in reality is the temperature fluctuations $\delta{T}/T_0$ in the CMB and the inhomogeneous distribution of galaxies. According to the standard cosmology, they are originated from the initial conditions. Being (believed to be) a Gaussian random field, which is in very good agreement with recent observations, what is important for $\delta{T}/T_0$ is their statistical properties on the whole observed CMB surface, rather than the (classical) evolution of $\delta{T}({\bm x})/T_0$ at a certain spatial location back to the moment of generation. Given that the CMB anisotropy spectrum can be written as
\begin{equation}
C_\ell^{TT} \sim \int (\text{transfer function}) \times P_\calR(k) \, ,
\end{equation}
where the transfer function is completely fixed by the background hot big bang cosmological parameters, the initial scalar power spectrum $P_\calR(k)$ is all that matters. For inflationary cosmology, $P_\calR(k)$ is identified as the power spectrum of the comoving curvature perturbation. Likewise, the $B$-mode polarization power spectrum is supposed to be originated from that of the primordial tensor perturbations $P_h(k)$:
\begin{equation}
C_\ell^{BB} \sim \int (\text{transfer function}) \times P_h(k) \, ,
\end{equation}
with a different transfer function from the one for the temperature fluctuations. So, what is important for the classical stochastic field such as the temperature fluctuations and $B$-mode polarization in the CMB is the statistical properties, rather than the individual evolution of each component. In turn, what is important for us in talking about the cosmological perturbations in the current context is -- the classicality of the original quantum cosmological perturbations is not on whether we can describe the solution in a manner closest to the classical solution, but on whether we can identify the quantum probability density of a certain state as the classical probability distribution, because we are not directly observing the primordial cosmological perturbations individually but are only indirectly constraining them via correlation functions. And such a state is claimed to be the squeezed state.

This indeed is how precisely the ``classicality'' is achieved by the squeezed state. Through time evolution, the uncertainty of measuring the field operators and their conjugate momenta is rather exponentially increasing: $\Delta\phi\Delta\pi_\phi \propto e^{2t}$. That is, the probability distribution spreads over in both $\phi$ and $\pi_\phi$. However, at the same time, the expectation values $\langle \phi \pi_\phi \rangle$ and $\langle \pi_\phi \phi \rangle$ in terms of the squeezed state tend to converge to the exponentially large value and non-commutativity between $\phi$ and $\pi_\phi$ becomes suppressed~\cite{Albrecht:1992kf}.
Since it is a good approximation to assign $\phi$ and $\pi_\phi$ to certain values simultaneously, the probability distribution is a function of a set of values $(\phi,\pi_\phi)$ called the Wigner function which is interpreted as a classical probability distribution.

Once we identify the pointer bases as the set of the squeezed states like $U_{0,\calS}|0\rangle_\calS$ given by \eqref{eq:matrixbases}, the calculations are extremely straightforward. As we may read from \eqref{eq:schematic-Lindblad-matrix}, the Lindblad terms are outer products of various basis vectors \eqref{eq:matrixbases} along with time- and momentum-dependent coefficients. To extract the matrix elements of the Lindblad terms, say, $d\rho_\text{red}/d\tau|_{ab}$ with $a$ and $b$ being the indices of the basis vectors, from \eqref{eq:schematic-Lindblad-matrix} we simply compute
\begin{equation}
\frac{d\rho_\text{red}}{d\tau} \bigg|_{ab} 
\equiv \left\langle a \left| \frac{d\rho_\text{red}}{d\tau} \right| b \right\rangle
= 
\rho_{mn} \big\langle a \big|
U_{0,\calS} a_{1,\calS}a_{2,\calS}\cdots a_{m,\calS} |0\rangle_\calS
\langle0|_\calS a_{1',\calS}a_{2',\calS}\cdots a_{n,\calS} U_{0,\calS}^\dag
\big| b \big\rangle 
\, .
\end{equation}
Then the reduced density matrix equation is described by a $7\times7$ matrix, with each row and column being distinguished by the number of creation (or annihilation) operators acting on the vacuum state. After straight calculations and considering the contributions of the products of polarization tensors, we end up with the following evolution equation of the reduced density matrix:
\begin{equation}
\label{eq:matrix-schematic2}
\frac{d\rho_\text{red}}{d\tau} =
-\frac{1}{2}
\begin{pmatrix}
\mathfrak{E}_{00} & 0 & \mathfrak{E}_{02} & 0 & 0 & 0 & 0
\\
0 & \mathfrak{E}_{11} & 0 & 0 & 0 & 0 & 0
\\
\mathfrak{E}_{20} & 0 & 0 & 0 & 0 & 0 & 0
\\
0 & 0 & 0 & 0 & 0 & 0 & 0 
\\
0 & 0 & 0 & 0 & 0 & 0 & 0
\\
0 & 0 & 0 & 0 & 0 & 0 & 0
\\
0 & 0 & 0 & 0 & 0 & 0 & 0
\end{pmatrix}
+ h.c.
\, ,
\end{equation}
where 
\begin{align}
\label{eq:matrix-00}
\mathfrak{E}_{00} 
& =
\frac{2}{(2\pi)^3} \delta^{(3)}({\bm q}) \sum_{\lambda_i}
\int_\calS d^3k_1 \int_\calE d^3k_2 
\nonumber\\
& \qquad \times
\Big[ h_0({\bm k}_2,{\bm k}_1,-{\bm k}_{12}+{\bm q}) 
+ h_0({\bm k}_1,{\bm k}_2,-{\bm k}_{12}+{\bm q}) 
+ h_0({\bm k}_1,-{\bm k}_{12}+{\bm q},{\bm k}_2) \Big]
\nonumber\\
& \qquad \times
\Big[
g_0^*(-{\bm k}_1,-{\bm k}_2,{\bm k}_{12}-{\bm q})
+ \big( \text{5 perm in arranging } 
-{\bm k}_1,-{\bm k}_2,{\bm k}_{12}-{\bm q} \text{ in } g_0^* \big)
\Big] 
\, ,
\\
\label{eq:matrix-02}
\mathfrak{E}_{02} 
& =
\delta^{(3)}({\bm q}_{ab})
\int_\calE
d^3k_1 
\sum_{\lambda_i}
3 \Big[ h_0({\bm k}_1,-{\bm k}_1+{\bm q}_b,{\bm q}_a) 
+ h_0({\bm k}_1,{\bm q}_a,-{\bm k}_1+{\bm q}_b) 
+ h_0({\bm q}_a,{\bm k}_1,-{\bm k}_1+{\bm q}_b) \Big]
\nonumber\\
& \quad \times
\Big[ g_1^*({\bm q}_b,-{\bm k}_1,{\bm k}_1-{\bm q}_b) 
+ g_1^*({\bm q}_b,{\bm k}_1-{\bm q}_b,-{\bm k}_1) \Big]
+
({\bm q}_a \leftrightarrow {\bm q}_b)
\, ,
\\
\label{eq:matrix-11}
\mathfrak{E}_{11} 
& = 
- 2 \delta^{(3)}({\bm q}_{ab})
\int_\calE
d^3k_1 
\sum_{\lambda_i}
\Big[
h_0({\bm q}_a,{\bm k}_1,-{\bm k}_1+{\bm q}_b) 
+ h_0({\bm k}_1,{\bm q}_a,-{\bm k}_1+{\bm q}_b) 
+ h_0({\bm k}_1,-{\bm k}_1+{\bm q}_b,{\bm q}_a) \Big]
\nonumber\\
& \quad \times
\Big[ g_0^*({\bm q}_b,-{\bm k}_1,{\bm k}_1-{\bm q}_b)
+ \big( \text{5 perm in arranging } {\bm q}_b,-{\bm k}_1,{\bm k}_1-{\bm q}_b
\text{ in } g_0^* \big)
\Big]
\, ,
\\
\label{eq:matrix-20}
\mathfrak{E}_{20} 
& = 
3 \delta^{(3)}({\bm q}_{ab})
\int_\calE
d^3k_1 
\sum_{\lambda_i}
h_1({\bm q}_a,{\bm k}_1,-{\bm k}_1+{\bm q}_b) 
\nonumber\\
& \quad \times
\Big[ g_0^*({\bm q}_b,-{\bm k}_1,{\bm k}_1-{\bm q}_b)
+ \big( \text{5 perm in arranging } {\bm q}_b,-{\bm k}_1,{\bm k}_1-{\bm q}_b
\text{ in } g_0^* \big) \Big]
+ ({\bm q}_a \leftrightarrow {\bm q}_b)
\, ,
\end{align}
with ${\bm k}_{12} = {\bm k}_1+{\bm k}_2$ and likewise ${\bm q}_{ab} = {\bm q}_a+{\bm q}_b$, with ${\bm q}_a$ and ${\bm q}_b$ being the momenta of external in- and out-states. Note that in all contributions the configuration is the product of two identical momentum triangles with two momenta in the environment sector and the remaining one in the system sector, $(\calE\calE\calS)\times(\calE\calE\calS)$.

\section{Evolution of reduced density matrix}
\label{sec:solution}
\setcounter{equation}{0}

\subsection{Solution of reduced density matrix}

To find explicitly the coefficients of $d\rho_\text{red}/d\tau$, we first note that since $e_{ij}^\lambda$ is symmetric under $i \leftrightarrow j$, the trace of any permutation is the same: 
\begin{equation}
{\rm Tr} \Big[ e^1(\hat{\bm k}_1)e^2(\hat{\bm k}_2)e^3(\hat{\bm k}_3) \Big]
= \text{the same for any order of } e^1(\hat{\bm k}_1), e^2(\hat{\bm k}_2), e^3(\hat{\bm k}_3)
\, .
\end{equation}
Moreover, from the cosine law
\begin{equation}
{\bm k}_1\cdot{\bm k}_2 + {\bm k}_2\cdot{\bm k}_3 + {\bm k}_3\cdot{\bm k}_1 
= -\frac{1}{2} \big( k_1^2 + k_2^2 + k_3^2 \big) 
\, ,
\end{equation}
we can write $h_0$ as
\begin{equation}
h_0({\bm k}_1,{\bm k}_2,{\bm k}_3;\tau)
=
-i \frac{H}{\mpl} \frac{e^{-ik_{123}\tau}}{(k_1k_2k_3)^{3/2}} \frac{1}{\tau^4}
\Big[ 
- 2 e^{\lambda_1}_{ij}({\bm k}_1)e^{\lambda_2}_{jk}({\bm k}_2)
e^{\lambda_3}_{ki}({\bm k}_3)
+
\text{higher power in }\tau 
\Big]
\, ,
\end{equation}
where $k_{123} = k_1+k_2+k_3$. Here, we have arranged the terms following the power of $\tau$, as eventually we will be interested in the moment near the end of inflation, $\tau \to 0$. Notice that $h_0$ is symmetric under the exchange of momenta. Integrating over $\tau$, we find
\begin{equation}
\label{eq:g0-int}
g_0({\bm k}_1,{\bm k}_2,{\bm k}_3;\tau) 
= 
\int_{\tau_0}^\tau d\tau' h_0(\tau')
=-i \frac{H}{\mpl} \frac{e^{-ik_{123}\tau}}{(k_1k_2k_3)^{3/2}} 
\bigg[
\frac{2}{3\tau^3} e^{\lambda_1}_{ij}({\bm k}_1)
e^{\lambda_2}_{jk}({\bm k}_2) e^{\lambda_3}_{ki}({\bm k}_3)
+
\text{higher power in }\tau  
\bigg]
\, ,
\end{equation}
where we have used ${\rm Ei}(-ikx) \underset{x\to-\infty}{\longrightarrow} -i\pi$ and dropped such contributions since they are, in the limit $\tau \to 0$, sub-dominant compared to the terms with inverse powers of $\tau$. Again, as $h_0$, $g_0$ is also symmetric under the exchange of momenta. Thus,
\begin{equation}
h_0g_0^* 
=
- \frac{H^2}{\mpl^2} \frac{1}{(k_1k_2k_3)^3} \frac{4}{3\tau^7}
\Big[ e^{\lambda_1}_{ij}({\bm k}_1) e^{\lambda_2}_{jk}({\bm k}_2) e^{\lambda_3}_{ki}(-{\bm k}_{12}) \Big] 
\Big[ e^{\lambda_1}_{i'j'}(-{\bm k}_1) e^{\lambda_2}_{j'k'}(-{\bm k}_2) e^{\lambda_3}_{k'i'}({\bm k}_{12}) \Big]
+ \cdots 
\, .
\end{equation}

To proceed further, we need to compute the product of the polarization tensors. From the homogeneity and isotropy of the background, without loss of generality by arranging ${\bm k}_1$ and ${\bm k}_2$ in such a way that ${\bm k}_1$ is along the $z$-direction and ${\bm k}_2$ is confined in the $xz$-plane with the polar angle $\theta$, i.e. 
\begin{equation}
\begin{split}
{\bm k}_1 & = (0,0,k_1) \, ,
\\
{\bm k}_2 & = (k_2\sin\theta,0,k_2\cos\theta) \, ,
\end{split}
\end{equation}
we can explicitly compute the polarization tensors for each vector from \eqref{eq:ptensor-p} and \eqref{eq:ptensor-c}. Summing over all possible combinations of polarization indices we find
\begin{align}
\label{eq:polsum-explicit}
&
\sum_{\lambda_i}
\Big[ e^{\lambda_1}_{ij}({\bm k}_1) e^{\lambda_2}_{jk}({\bm k}_2) 
e^{\lambda_3}_{ki}(-{\bm k}_{12}) \Big] 
\Big[ e^{\lambda_1}_{i'j'}(-{\bm k}_1) e^{\lambda_2}_{j'k'}(-{\bm k}_2) 
e^{\lambda_3}_{k'i'}({\bm k}_{12}) \Big]
\nonumber\\
& =
\frac{\Big[ 4k_1^4 + 4k_2^4 + 11k_1^2k_2^2 
+ 8 k_1k_2 \big( k_1^2 + k_2^2 \big) \cos\theta
+ k_1^2k_2^2 \big( 2\cos^2\theta - 1 \big) \Big] 
\sin^4\theta}{2 \big( k_1^2 + k_2^2 + 2k_1k_2\cos\theta \big)^2}
\, .
\end{align}
Then, $\mathfrak{E}_{00}$ becomes
\begin{align}
\label{eq:2integral-1}
\mathfrak{E}_{00} 
& =
\frac{2}{(2\pi)^3} \delta^{(3)}({\bm q}) 
\int_\calS d^3k_1 \int_\calE d^3k_2 \sum_{\lambda_i}
h_0({\bm k}_1,{\bm k}_2,-{\bm k}_{12}+{\bm q})
\times 6g_0^*(-{\bm k}_1,-{\bm k}_2,{\bm k}_{12}-{\bm q})
\nonumber\\
& =
\frac{2}{(2\pi)^3} \delta^{(3)}({\bm q}) 
\frac{-H^2}{\mpl^2} \frac{4}{3\tau^7} 8\pi^2 \frac{1}{\calH^3}
\int_0^{1} d\kappa_1 \int_{1}^\infty d\kappa_2 \int_{-1}^1 d(\cos\theta)
\frac{1}{\kappa_1\kappa_2 \big( \kappa_1^2 + \kappa_2^2 
+ 2\kappa_1\kappa_2\cos\theta \big)^{3/2}} 
\nonumber\\
& \hspace{7em}
\times
\frac{\big[ 4\kappa_1^4 + 4\kappa_2^4 + 11\kappa_1^2\kappa_2^2 + 8\kappa_1\kappa_2 \big( \kappa_1^2 + \kappa_2^2 \big) \cos\theta
+ \kappa_1^2\kappa_2^2 \big( 2\cos^2\theta - 1 \big) \big] \sin^4\theta}{2 \big( \kappa_1^2 + \kappa_2^2 + 2\kappa_1\kappa_2\cos\theta \big)^2} 
\nonumber\\
& \equiv
\frac{2}{(2\pi)^3} \delta^{(3)}({\bm q}) 
\frac{H^2}{\mpl^2} \frac{4}{3\tau^4} 8\pi^2
\calC_{\calS\calE} 
\, ,
\end{align}
where $\kappa_1 = k_1/\calH = -k_1\tau$ and $\kappa_2 = k_2/\calH = -k_2\tau$. Note that being in the system (environment) sector, the magnitude of $\kappa_1$ ($\kappa_2$) is bounded up (down) to the horizon scale, i.e. $k_1 (k_2) = \calH = aH$. In the last equality we have used $\calH = -1/\tau$, and we have defined the integrations over $\kappa_1$, $\kappa_2$ and $\cos\theta$ as a coefficient $\calC_{\calS\calE}$. Also notice that overall there would remain no dependence on ${\bm k}_1$ or ${\bm k}_2$, but only with weak dependence on the infrared cutoff that can be inferred by counting the power of $\kappa_1$ and $\kappa_2$. Indeed, we can perform the integral analytically to find $\calC_{\calS\calE}$ explicitly as
\begin{equation}
\label{eq:CSE}
\calC_{\calS\calE} 
=
\frac{5056}{4725} - \frac{32}{45}\log2 
- \frac{32}{45}\log\varepsilon
\approx 
0.577148 - \frac{32}{45}\log\varepsilon
\, ,
\end{equation}
where $\varepsilon \ll 1$ is the infrared cutoff for $\kappa_1$.

We can proceed in a similar manner to find $\mathfrak{E}_{11}$ to find, with ${\bm q}_a = -{\bm q}_b \equiv {\bm q}$,
\begin{align}
\label{eq:1integral-1}
\mathfrak{E}_{11}
& =
-2 \delta^{(3)}({\bm q}_{ab}) \int_\calE d^3k_1 \sum_{\lambda_i}
3 h_0({\bm q},{\bm k}_1,-{\bm k}_1-{\bm q}) \times 6 g_0^*(-{\bm q},-{\bm k}_1,{\bm k}_1+{\bm q})
\nonumber\\
& =
-2 \delta^{(3)}({\bm q}_{ab}) \delta_{\lambda_a \lambda_b} 18 \frac{H^2}{\mpl^2} \frac{4}{3\tau^4}
\frac{2\pi}{q^3} \calC_\calE
\, ,
\end{align}
where the coefficient $\calC_\calE$ is given by, with $\bar{q} \equiv q/\calH < 1$,
\begin{align}
\label{eq:CE}
\calC_\calE
& \equiv
\int_1^\infty d\kappa_1 \int_{-1}^1 d(\cos\theta)
\frac{1}{\kappa_1 \big( \bar{q}^2 + \kappa_1^2 + 2\bar{q}\kappa_1\cos\theta \big)^{3/2} }
\nonumber\\
& \quad
\times 
\frac{\big[ 4\bar{q}^4 + 4\kappa_1^4 + 11\bar{q}^2\kappa_1^2 + 8\bar{q}\kappa_1 
\big( \bar{q}^2 + \kappa_1^2 \big) \cos\theta + \bar{q}^2\kappa_1^2 \cos(2\theta) \big] \sin^4\theta}
{2 \big( \bar{q}^2 + \kappa_1^2 + 2\bar{q}\kappa_1\cos\theta \big)^2}
\nonumber\\
& =
\frac{16}{525\bar{q}^3}
\bigg[ 8\bar{q}^5 - 70\bar{q} 
+ 35\log \bigg( \frac{1+\bar{q}}{1-\bar{q}} \bigg)
\bigg]
\, .
\end{align}
Note that $\calC_{\calS\calE}$ and $\calC_\calE$ are related by
\begin{equation}
\label{eq:CSE-CE}
\calC_{\calS\calE}
=
\int_{\varepsilon\ll1}^1 \bar{q}^2 d\bar{q} 
\frac{\calC_\calE}{\bar{q}^3}
\, ,
\end{equation}
as should be the case to maintain the traceless property of \eqref{eq:matrix-schematic2}.

We pause to comment on the infrared logarithmic divergence in $\calC_{\calS\calE}$. Even though $\mathfrak{E}_{00}$ is divergent as $\varepsilon \to 0$, the tracelessness of the Lindblad terms, or equivalently unit trace of the density matrix, guarantees that the divergence $\mathfrak{E}_{00}$ is canceled by that in $\mathfrak{E}_{11}$ as we can see from \eqref{eq:CSE-CE}. Indeed, when the external momentum ${\bm q}$ involved in $\mathfrak{E}_{11}$ is extremely soft, our observational apparatus does not distinguish $\mathfrak{E}_{11}$ from $\mathfrak{E}_{00}$. That is, if the momentum involved in the one-particle to one-particle process represented by $\mathfrak{E}_{11}$ is too small well below threshold resolution to be detected by any means, it is observationally identical to the process involving no particle excitation, $\mathfrak{E}_{00}$. In that sense, the infrared cutoff $\varepsilon$ reveals our limit of detecting the soft graviton, which suggests that we need to sum up $\mathfrak{E}_{00}$ and $\mathfrak{E}_{11}$ with $q < \varepsilon$ inclusively to obtain the probability of no effective detection of graviton at initial and final states. This is exactly how we obtain the finite infrared amplitude of the process in quantum electrodynamics~\cite{Bloch:1937pw}, even though the probability in this case is quantum mechanical whereas that in our density matrix is classical.

Finally, $\mathfrak{E}_{20}$ component is given by 
\begin{align}
\label{eq:1integral-2}
\mathfrak{E}_{20}
& =
3 \delta^{(3)}({\bm q}_{ab}) \int_\calE d^3k_1 \sum_{\lambda_i}
 h_0({\bm q},{\bm k}_1,-{\bm k}_1-{\bm q}) \times 6 g_0^*(-{\bm q},-{\bm k}_1,{\bm k}_1+{\bm q}) \times 2
\nonumber\\
& =
-3 \delta^{(3)}({\bm q}_{ab}) \delta_{\lambda_a \lambda_b} 12 \frac{H^2}{\mpl^2} \frac{4}{3\tau^4}
\frac{2\pi}{q^3} e^{2 i q \tau} \calC_\calE 
=
\mathfrak{E}_{02}^*
\, ,
\end{align}
which is just $\mathfrak{E}_{11}$ multiplied by $e^{2 i q \tau}$. Since $\mathfrak{E}_{11}$ or $\mathfrak{E}_{20}$ correspond to the matrix element with two external legs, the exponential factor $e^{-i q \tau}$ ($e^{i q \tau}$) can be understood as the phase factor of the graviton wave function for the in- (out-) state. So the phase factors cancel each other in $\mathfrak{E}_{11}$, meanwhile they are squared in $\mathfrak{E}_{20}$ and $\mathfrak{E}_{02}$. Thus, finally, the evolution equation of the reduced density matrix is given by
\begin{equation}
\frac{d\rho_\text{red}}{d\tau}
=
- \begin{pmatrix}
\mathfrak{E}_{00} & 0 & \mathfrak{E}_{02} 
& \vline &
\\
0 & \mathfrak{E}_{11} & 0 
& \vline & 0_{3\times4}
\\
\mathfrak{E}_{20} & 0 & 0
& \vline &
\\
\hline
& 0_{4\times3} & 
& \vline & 0_{4\times4}
\end{pmatrix}
\, ,
\end{equation}
with $\mathfrak{E}_{00}$, $\mathfrak{E}_{11}$ and $\mathfrak{E}_{20} = \mathfrak{E}_{02}^*$ given respectively by \eqref{eq:2integral-1}, \eqref{eq:1integral-1} and \eqref{eq:1integral-2}.

\subsection{Decoherence rate}

Now, we can discuss the non-unitary effect in the reduced density matrix. First of all, the $00$ element of the reduced density matrix $\rho_{\rm red}$ is given by (see Appendix~\ref{app:matrix})
\begin{equation}
\label{eq:rho00}
\rho_{\rm red}|_{00}
=
1 - \frac{2}{(2\pi)^3} \delta^{(3)}({\bm q}) \frac{H^2}{\mpl^2}
\frac{4}{9|\tau|^3} 8\pi^2 \calC_{\calS\calE}
\, .
\end{equation}
This corresponds to the otherwise unitary evolution of the squeezed state, $U_{0,\calS}|0\rangle_\calS$, i.e. the evolution from the initial Bunch-Davies system vacuum with which we have started. That is, if we take the interaction Hamiltonian $H_\text{int}$ into account,

\begin{enumerate}

\item $H_\text{int}$ on one hand generates, as shown in \eqref{eq:Hint-onlysystem}, the unitary evolution in the system sector to produce many non-zero elements of the density matrix. We can calculate them explicitly by following similar steps presented in Appendix~\ref{app:fullrho} but here we do not show them as they are sub-leading unitary evolution effects.

\item On the other hand, $H_\text{int}$ also generates the non-unitary evolution through the Lindblad terms, reducing the probability to maintain the pure squeezed state from unity. That means, the reduced matrix element is no longer restricted to the $00$ element, but spreads out to produce other reduced density matrix elements, which are interpreted as the generation of classical probabilities for corresponding processes. Still, the trace of the reduced density matrix is unity, preserving the total classical probability. In our case, $\rho_{\rm red}|_{11}$ is produced and ${\rm Tr} \big[ \rho_{\rm red}|_{00} + \rho_{\rm red}|_{11} ] = 1$, where the trace is taken over the momentum space and polarization, guarantees this fact. At the same time, $\rho_{\rm red}|_{20}$ and $\rho_{\rm red}|_{02}$ are also produced since they are equivalent to one-particle to one-particle process: just exchanges of the in and out states. The non-unitary evolution is of second order in $H_\text{int}$ as shown in \eqref{eq:rho00}. Hence, our calculation is valid provided that we can treat $H_\text{int}$ perturbatively.

\end{enumerate}

We can further read from \eqref{eq:rho00} the rate that the pure state evolves into the mixed state through the non-unitary Lindblad terms -- the {\it decoherence rate} per unit volume for the whole system modes. By exponentiating \eqref{eq:rho00}, and noting that ${\bm q}$ is an almost vanishing momentum so $(2\pi)^3 \delta^{(3)}({\bm q})$ corresponds to the comoving volume of three-dimensional space $V$ (see Footnote~\ref{footnote}), we can write  $\rho_\text{red}|_{00} = \exp(-V \int \Gamma d\tau)$ where, from \eqref{eq:2integral-1},
\begin{equation}
\label{eq:rate}
\Gamma 
= 
\frac{2}{(2\pi)^{3\cdot2}} \frac{H^2}{\mpl^2} 
\frac{4}{3\tau^4} 8\pi^2\calC_{\calS\calE}
=
\frac{\Delta_\calR^2}{(2\pi)^{2}} 
\frac{2}{3\tau^4} r\calC_{\calS\calE}
\, .
\end{equation}
Here $r \equiv \Delta_h^2/\Delta_\calR^2$ is the tensor-to-scalar ratio and
\begin{equation}
\Delta_\calR^2 = \frac{1}{2\epsilon\mpl^2} \bigg( \frac{H}{2\pi} \bigg)^2
\quad \text{and} \quad
\Delta_h^2 = \frac{8}{\mpl^2} \bigg( \frac{H}{2\pi} \bigg)^2
\end{equation}
are the amplitudes of the scalar and tensor perturbations respectively. The decoherence rate \eqref{eq:rate} grows as the physical volume $a^3$, but is suppressed by $r\Delta_\calR^2$, or equivalently $H^2/\mpl^2$. That is, the suppression of $\Gamma$ is because the inflationary scale is well below the full quantum gravity regime, {\it irrespective} of the inflationary dynamics. Further, for an interval of $\Delta\tau$, we find the change in $\rho_\text{red}|_{00}$ is
\begin{equation}
\exp \bigg[ - \frac{\Delta_\calR^2}{(2\pi)^{2}}  
\frac{2}{9}e^{3\Delta{N}} r\calC_{\calS\calE} \bigg]
\, ,
\end{equation}
where $\Delta{N}$ is the number of $e$-folds elapsed during $\Delta\tau$. Here, we have absorbed the factor $VH^{-3}$ into the normalization of the scale factor so that only the difference between $a(\tau)$ and $a(\tau+\Delta\tau)$, viz. $\Delta{N}$, is highlighted. In other words, we need
\begin{equation}
\label{eq:Ndec}
\Delta{N}_\text{dec} 
\approx 
\frac{1}{3} \log \bigg[ \frac{(2\pi)^2}{\Delta_\calR^2} \frac{9}{2} 
\big( r\calC_{\calS\calE} \big)^{-1} \bigg] 
\approx
8.38689 - \frac{1}{3}\log \big( r\calC_{\calS\calE} \big)
\end{equation}
for $\rho_\text{red}|_{00}$ to change by $e^{-1}$, i.e. for decoherence to occur, where we have taken the central value from the Planck 2018 result $\Delta_\calR^2 \approx 2.101 \times 10^{-9}$~\cite{Akrami:2018odb}. For a wide range of $\calC_{\calS\calE}$ and $r$, typically $5 \lesssim \Delta{N}_\text{dec} \lesssim 10$.

Finally, notice that \eqref{eq:1integral-1} can be compared to the decoherence rate for the curvature perturbation $\calR$ representing the reduced matrix element $\langle \calR_{\bm q} | \rho_{\rm red} | \calR_{\bm q} \rangle$ given by~\cite{Nelson:2016kjm} 
\begin{equation}
\label{eq:deccurv}
\langle \calR_{\bm q} | \rho_{\rm red} | \calR_{\bm q} \rangle
\approx
1 - \frac{(\epsilon+\eta)^2}{288}
\Delta_\calR^2 \frac{1}{|q\tau|^3} 
\, ,
\end{equation}
with $\eta \equiv \dot\epsilon/(H\epsilon)$. Here, the momentum configuration is the squeezed one where the system momentum ${\bm q}$ is much smaller than the environment one. The unity in the right-hand side indicates that the initial pure state is implemented by $|\calR_{\bm q}\rangle$: without interaction Hamiltonian, we expect $\langle \calR_{\bm q} | \rho_{\rm red} | \calR_{\bm q} \rangle=1$. In addition, we find that it has the similar structure as \eqref{eq:1integral-1}. That means there should be similar features in the density matrix -- for example the $q^{-3}$ dependence implies that taking the trace of the density matrix over one particle states results in a logarithmic infrared divergence, which should be cancelled by other diagonal density matrix elements, e.g. $\rho_{\rm red}|_{00}$. On the other hand, there are two differences: 1) we have an additional delta function $\delta^{(3)}({\bm q}_{ab})$ representing that in- and out-state have the same momentum. This is a result of the contraction between creation (annihilation) operator in $H_\text{int}$ of $L_1$ and annihilation (creation) one in Hamiltonians of $L_2$, and 2) the appearance of the slow-roll parameters as expected. This reflects that the Goldstone boson nature of the curvature perturbation resulting from the spontaneous breaking of dS isometry~\cite{Gong:2017wgx}. As a result, the number of $e$-folds for decoherence to occur has an additional slow-roll suppression to give $\Delta{N}_\text{dec}$ typically two times bigger than \eqref{eq:Ndec}, $\Delta N_\text{dec} \gtrsim 10$.

\section{Conclusions}
\label{sec:conclusion}

In this article, we have considered in the dS background the time evolution of the squeezed state for tensor perturbations under the influence of non-linear interaction Hamiltonian. From this, we have shown how tracing out the unobservable sub-horizon modes converts the pure state, i.e. the initial squeezed state, into the mixed state through the Lindblad operators from the non-linear cubic interactions, in which super- (system) and sub- (environment) horizon modes interact with each other. Since the mixed state is interpreted as an ensemble of the pure states with classical probabilities represented by the density matrix elements, the process we have described is interpreted as a quantum-to-classical transition that breaks the unitarity. This needs to be distinguished from the traditional quantum-to-classical transition accomplished by the squeezed state. That is, the squeezed state can be understood as the time evolution of the vacuum state by the unitary operator, coming from the quadratic sector in which the interaction between super- and sub-horizon modes is not taken into account. In this case, the classicality is coming from suppression of non-commutativity between the field variables and their conjugate momenta.

The probability of maintaining the squeezed state for the tensor perturbations is given by~\eqref{eq:rho00}. The fact that the reduced density matrix has the unit trace is reflected in the generation of other matrix elements, say, $11$ and equivalent $20$ and $02$ elements in our case. While we have found that only the processes in which two particles are involved -- one- to one-particle, zero- to two-particle and two- to zero-particle -- considering higher order in the interaction Hamiltonian should excite the processes with more particles involved. Furthermore, the rate of decoherence, the non-unitary quantum-to-classical transition, can be computed explicitly. This can be read from the rate of change in the $00$ element of the reduced density matrix element and is given by \eqref{eq:rate}. Typically, it takes $5 \lesssim \Delta N_\text{dec} \lesssim 10$ $e$-folds for decoherence to occur rapidly. This delay is more or less half of that for the curvature perturbation, because it accompanies one more slow-roll suppression due to its Goldstone boson nature, so that the decoherence occurs slower accordingly.

\subsection*{Acknowledgements}

We thank Toshifumi Noumi, Gary Shiu, Jiro Soda, Yuko Urakawa and Jaiyul Yoo for discussions while this work was under progress.
JG is supported in part by the Basic Science Research Program through the National Research Foundation of Korea Research Grant (2016R1D1A1B03930408). JG also acknowledges the Korea-Japan Basic Scientific Cooperation Program supported by the National Research Foundation of Korea and the Japan Society for the Promotion of Science (2018K2A9A2A08000127).

\newpage

\appendix

\renewcommand{\theequation}{\Alph{section}.\arabic{equation}}

\section{Quadratic evolution of tensor perturbations}
\label{app:quadratic}
\setcounter{equation}{0}

\subsection{Solutions of operators}

Considering the spatial metric as\footnote{We are ignoring pure scalar and vector perturbations. But at quadratic level all perturbations of different kinds are decoupled so it does not matter.}
\begin{equation}
g_{ij} = a^2(t) \big( \delta_{ij} + h_{ij} \big) 
\, ,
\end{equation}
where $h_{ij}$ is the pure tensor perturbations with $h^i_i = 0$ and $\partial_i h^i_j = 0$, the quadratic action of $h_{ij}$ becomes
\begin{equation}
S_2^{(t)} = \int d\tau d^3x \frac{a^2\mpl^2}{8} 
\Big[ {h_{ij}'}^2 - (\nabla{h}_{ij})^2 \Big] \, .
\end{equation}
Here, $d\tau \equiv dt/a$ is the conformal time, and the indices of $h_{ij}$ are raised and lowered by $\delta_{ij}$ so we do not sharply distinguish the upper and lower indices of $h_{ij}$, e.g. ${{h}_{ij}'}^2$ should be understood as ${{h}^{ij}}{h}_{ij}'$. Since there are two physical degrees of freedom for $h_{ij}$, we introduce the polarization tensor $e_{ij}(\lambda)$, with $\lambda$ being the polarization index, which satisfies\footnote{In fact, this normalization is arbitrary. It is, along with the normalization of the mode function \eqref{eq:canonicaltensor}, set to give the standard consistency relation $r=16\epsilon$ for single-field inflation.}
\begin{equation}
\label{eq:pol-orthogonal}
e_{ij}(\lambda)e_{ij}(\lambda') = 2\delta_{\lambda\lambda'} 
\, ,
\end{equation}
so that we can write
\begin{equation}
\label{eq:polarization-decompose}
h_{ij} = \sum_{\lambda=1}^2 h_\lambda e_{ij}(\lambda) 
\, .
\end{equation}
For canonical normalization, we introduce
\begin{equation}
\label{eq:canonicaltensor}
v_\lambda \equiv \frac{a\mpl}{\sqrt{2}} h_\lambda 
\, ,
\end{equation}
so that the quadratic action becomes
\begin{equation}
\label{eq:S2t}
S_2^{(t)} 
= \sum_\lambda \int d\tau d^3x \frac{1}{2} \left[
{v_\lambda'}^2 - 2\frac{a'}{a}v_\lambda'v_\lambda - (\nabla v_\lambda)^2 + \left( \frac{a'}{a} \right)^2 v_\lambda^2
\right] 
\, .
\end{equation}
Thus there are two copies of the identical action of a canonically normalized scalar field $v_\lambda$ for each polarization state. So from now on we just consider only one polarization state and drop the subscript $\lambda$, since for the other polarization we can follow precisely the same steps to describe the physical evolution of the system.

Now from \eqref{eq:S2t} we define the conjugate momentum $\pi$ as usual:
\begin{equation}
\label{eq:conj-momentum}
\pi \equiv \frac{\partial\calL(v,v')}{\partial{v'}} 
= v' - \frac{a'}{a}v \, .
\end{equation}
Then in terms of the Fourier mode as
\begin{equation}
v({\bm x}) = 
\int \frac{d^3k}{(2\pi)^3} e^{i{\bm k}\cdot{\bm x}} v_{\bm k} 
\, ,
\end{equation}
and the same for $\pi({\bm x})$, the Hamiltonian $H = \int d^3x \big( \pi v' - \calL \big)$ becomes 
\begin{equation}
\label{eq:H}
H
= 
\frac{1}{2} \int \frac{d^3k}{(2\pi)^3} 
\left[ \pi_{\bm k}\pi_{-{\bm k}} + k^2 v_{\bm k} v_{-{\bm k}}
+ \frac{a'}{a} \big( \pi_{\bm k} v_{-{\bm k}} 
+ v_{\bm k} \pi_{-{\bm k}} \big) \right] 
\, .
\end{equation}

Since we are interested in the evolution of the state, we may work in the Schr\"odinger picture where the time evolution is taken by the state and the operators remain fixed. For the study of cosmological perturbations usually the Heisenberg picture is taken where the operators are time evolving while the state is fixed, usually the vacuum. But in fact, it does not matter sharply in which picture we are working, since the physical results should be the same, especially at quadratic level. To proceed, we first promote the Fourier modes $\pi_{\bm k}$ and $v_{\bm k}$ to the time-dependent operators $\hat\pi_{\bm k}(\tau)$ and $\hat{v}_{\bm k}(\tau)$. In terms of the creation and annihilation operators $\hat\pi_{\bm k}(\tau)$ and $\hat{v}_{\bm k}(\tau)$ are given by
\begin{equation}
\label{eq:variables}
\begin{split}
\hat\pi_{\bm k}(\tau) & = a_{\bm k}(\tau)u_k + a_{-{\bm k}}^\dag(\tau) u_k^* \, ,
\\
\hat{v}_{\bm k}(\tau) & = a_{\bm k}(\tau)v_k + a_{-{\bm k}}^\dag(\tau) v_k^* \, .
\end{split}
\end{equation}
Note that with the operators being time-dependent, we are in the Heisenberg picture. One important difference from the conventional approach in the Heisenberg picture is that, the time dependence is given to the creation and annihilation operators, $a_{\bm k} = a_{\bm k}(\tau)$ and $a_{\bm k}^\dag = a_{\bm k}^\dag(\tau)$. Instead, the mode functions are set up at an initial time, $\tau=\tau_0$. Meanwhile, in the standard approach the creation and annihilation operators are defined initially and the mode functions carry the time dependence. This difference from the standard approach is to make the transfer from the Heisenberg picture to the Schr\"odinger picture more easily.

Being canonical, we have the following canonical commutation relations:
\begin{equation}
\label{eq:commutation}
\begin{split}
\left[ \hat{v}(\tau,{\bm x}), \hat\pi(\tau,{\bm y}) \right] & = i\delta^{(3)}({\bm x}-{\bm y}) \, ,
\\
\left[ a_{\bm k}(\tau), a_{\bm q}^\dag(\tau) \right] & = (2\pi)^3 \delta^{(3)}({\bm k}-{\bm q}) \, ,
\quad \text{otherwise zero} \, .
\end{split}
\end{equation}
Notice that these relations are non-zero only for the same polarization. If we revive the polarization index, we have $\left[a_{\bm k}^{\lambda}(\tau), a_{\bm q}^{\lambda'\dag}(\tau)\right] = (2\pi)^3 \delta_{\lambda\lambda'} \delta^{(3)}({\bm k}-{\bm q})$. Using the decompositions \eqref{eq:variables} and the commutation relations impose that, at a initial time $\tau_0$, the mode functions $v_k$ and $u_k$ satisfy
\begin{equation}
\label{eq:Wronskian}
u_k^*v_k - u_kv_k^* = i \, .
\end{equation}

To set up the initial conditions at $\tau = \tau_0$, we assume that the modes are deep inside the horizon, or the expansion of the universe is neglected. That is, we may only consider the decoupled first two terms in \eqref{eq:H}, say $\widehat{H}_0$, to set up the initial conditions. Then, we can determine the \textit{initial} mode function solutions by demanding that the expectation value of the free Hamiltonian operator $\widehat{H}_0$ is minimized at $\tau = \tau_0$:
\begin{equation}
\label{eq:Hmin}
\left\langle \widehat{H}_0(\tau_0) \right\rangle 
= \frac{1}{2} \int \frac{d^3k}{(2\pi)^3} 
(2\pi)^3 \delta^{(3)}({\bm 0}) \Big[ |u_k|^2 + k^2|v_k|^2 \Big] 
\, ,
\end{equation}
where the expectation value is taken with respect to the initial vacuum state, which is assumed to match the standard vacuum state in Minkowski space as the expansion of the universe can be ignored. We can easily find the solutions:
\begin{equation}
u_k = -i \sqrt{\frac{k}{2}}
\quad \text{and} \quad
v_k = \frac{1}{\sqrt{2k}}
\, ,
\end{equation}
and the canonical variables \eqref{eq:variables} become
\begin{equation}
\label{eq:sol}
\begin{split}
\hat\pi_{\bm k}(\tau) & =  -i \sqrt{\frac{k}{2}} \Big[ a_{\bm k}(\tau) - a_{-{\bm k}}^\dag(\tau) \Big] \, ,
\\
\hat{v}_{\bm k}(\tau) & = \frac{ a_{\bm k}(\tau) + a_{-{\bm k}}^\dag(\tau) }{\sqrt{2k}} \, .
\end{split}
\end{equation}

Substituting the solutions \eqref{eq:sol} into \eqref{eq:H}, from the Hamiltonian equations for the operators we can find the following coupled differential equations that $a_{\bm k}(\tau)$ and $a_{-{\bm k}}^\dag(\tau)$ should satisfy:
\begin{equation}
\label{eq:eom-operators}
\begin{split}
a_{\bm k}'(\tau) & = -ika_{\bm k}(\tau) + \frac{a'}{a} a_{-{\bm k}}^\dag(\tau) \, ,
\\
{a_{-{\bm k}}^\dag}'(\tau) & = ika_{-{\bm k}}^\dag(\tau) + \frac{a'}{a} a_{\bm k}(\tau) \, .
\end{split}
\end{equation}
The general solution of these equations are given by the linear combination of the initial ones:
\begin{equation}
\label{eq:bogoliubov}
\begin{split}
a_{\bm k}(\tau) & = \alpha_k(\tau) a_{\bm k}(\tau_0) + \beta_k(\tau) a_{-{\bm k}}^\dag(\tau_0) \, ,
\\
a_{-{\bm k}}^\dag(\tau) & = \alpha_k^*(\tau) a_{-{\bm k}}^\dag(\tau_0) + \beta_k^*(\tau) a_{\bm k}(\tau_0) \, ,
\end{split}
\end{equation}
which is the so-called Bogoliubov transformation. The commutation relations \eqref{eq:commutation} lead to the constraint $\alpha_k(\tau)$ and $\beta_k(\tau)$ are always subject to:
\begin{equation}
\label{eq:constraint}
|\alpha_k(\tau)|^2 - |\beta_k(\tau)|^2 = 1 
\, ,
\end{equation}
and we can parametrize them in terms of the hyperbolic functions as
\begin{equation}
\label{eq:alpha-beta}
\begin{split}
\alpha_k & = e^{-i\theta_k}\cosh r_k \, ,
\\
\beta_k & = e^{i(\theta_k + 2\varphi_k)} \sinh r_k \, ,
\end{split}
\end{equation}
where without loss of generality we take the time-dependent functions $r_k(\tau)$, $\varphi_k(\tau)$ and $\theta_k(\tau)$ real. Assuming a perfect de Sitter background so that $a'/a = -1/\tau$, we can find analytically the solutions as~\cite{Polarski:1995jg,Albrecht:1992kf}
\begin{align}
\label{eq:rk-sol}
r_k & = \sinh^{-1} \left( \frac{1}{2k\tau} \right)
\, ,
\\
\label{eq:varphik-sol}
\varphi_k & = \frac{\pi}{4} - \frac{1}{2} \tan^{-1} 
\left( \frac{1}{2k\tau} \right) 
\, ,
\\
\label{eq:thetak-sol}
\theta_k & = k\tau + \tan^{-1}\left( \frac{1}{2k\tau} \right)
\, .
\end{align}

\subsection{Evolution of the vacuum state}

Having found the canonical variables in terms of the time-dependent creation and annihilation operators, now we can write the Hamiltonian density operator as
\begin{equation}
\label{eq:Hoperator}
\widehat{\calH}(\tau) 
= 
\frac{1}{2} \Bigg\{ k \Big[ (2\pi)^3\delta^{(3)}({\bm 0})
+ a_{\bm k}^\dag a_{\bm k} + a_{-{\bm k}}^\dag a_{-{\bm k}} \Big]
+ i\frac{a'}{a} \Big( -a_{\bm k}a_{-{\bm k}} 
+ a_{-{\bm k}}^\dag a_{\bm k}^\dag \Big) \Bigg\} \, .
\end{equation}
Here, we have changed  the order of $a_{\bm k}a_{\bm k}^\dag$ using the commutation relation of $a_{\bm k}$. The reason is because after all we are interested in the evolution of the initial vacuum state. So the annihilation operator on the right first annihilate the vacuum state, so that the operation of this part is to keep the vacuum state intact as we will see right below. For clarity, now let us concentrate on a single $k$-mode. This means now the Hamiltonian we need to consider is $\widehat{\calH}$, not necessarily $\widehat{H} = \int d^3k/(2\pi)^3 \widehat\calH$. Equivalently, we can interpret this as ``discretization'' of the momentum, $\int d^3k/(2\pi)^3 \to L^{-3} \sum_k$ where $L^3$ is the volume into which the mode of our interest permeates. With a given volume $L^3$, we may now isolate the volume from the canonical creation and annihilation operators \eqref{eq:commutation} in such a way that, since ${a}_{\bm k}$ has a mass dimension of $M^{-3/2}$,
\begin{equation}
\label{eq:rescaled-op}
a_{\bm k} \equiv L^{3/2} \hat{a}_{\bm k}
\end{equation}
and likewise for $a_{\bm k}^\dag$. That means, the new (dimensionless) operators $\hat{a}_{\bm k}$ and $\hat{a}_{\bm k}^\dag$ now satisfies\footnote{
\label{footnote}
In that sense, $(2\pi)^3 \delta^{(3)}({\bm k} = {\bm 0})$ in \eqref{eq:Hoperator} can be understood as the comoving volume into which the quantum field of our interest permeate. From
\begin{equation*}
\int d^3x e^{i{\bm k}\cdot{\bm x}} = (2\pi)^3 \delta^{(3)}({\bm k}) \, ,
\end{equation*}
setting ${\bm k} = {\bm 0}$ tells us
\begin{equation*}
(2\pi)^3 \delta^{(3)}({\bm 0}) = \int d^3x = \text{volume} \, .
\end{equation*}
This corresponds to ``1'' in the commutation relation $[a,a^\dag] = 1$ for the one-dimensional harmonic oscillator.
} 
\begin{equation}
\left[ \hat{a}_{\bm k}, \hat{a}_{\bm k}^\dag \right] = 1 \, .
\end{equation}
Then the relation between $\widehat{H}$ and $\widehat{\calH}$ is also clear. From \eqref{eq:Hoperator}, we can see
\begin{equation}
\begin{split}
\widehat{\calH} & = L^3 \times \frac{1}{2} \Bigg\{ k \Big[ 1 + \hat{a}_{\bm k}^\dag\hat{a}_{\bm k}
+ \hat{a}_{-{\bm k}}^\dag\hat{a}_{-{\bm k}} \Big]
+ i\frac{a'}{a} \Big( -\hat{a}_{\bm k}\hat{a}_{-{\bm k}} + a_{-{\bm k}}^\dag\hat{a}_{\bm k}^\dag 
\Big) \Bigg\} \, ,
\\
\widehat{H} & = L^{-3} \sum_k \widehat{\calH} 
= \sum_k \frac{1}{2} \Bigg\{ k \Big[ 1 + \hat{a}_{\bm k}^\dag\hat{a}_{\bm k}
+ \hat{a}_{-{\bm k}}^\dag\hat{a}_{-{\bm k}} \Big]
+ i\frac{a'}{a} \Big( -\hat{a}_{\bm k}\hat{a}_{-{\bm k}} + a_{-{\bm k}}^\dag\hat{a}_{\bm k}^\dag 
\Big) \Bigg\} \, .
\end{split}
\end{equation}
So $\widehat{\calH}$ is the Hamiltonian density in the sense that it only accounts for the contribution of a single $k$-mode, but it includes the whole volume over which the quantum field is effective. Yet, $\widehat{H}$ is the Hamiltonian per unit volume, but the contribution of {\it all} $k$-modes is considered. So, in the conventional sense that density means something divided by the total volume, $\widehat{H}$ is Hamiltonian ``density'' because the effect of the total volume is singled out. To concentrate on a single $k$-mode [because we have already taken into account all delta functions in the commutation relations \eqref{eq:commutation}] let us write $\widehat{H} = \sum_k \widehat{\calH}_k$ where
\begin{equation}
\label{eq:Hoperator2}
\widehat{\calH}_k(\tau) 
= \frac{1}{2} \Bigg[ k \Big( 1 + \hat{a}_{\bm k}^\dag \hat{a}_{\bm k} 
+ \hat{a}_{-{\bm k}}^\dag \hat{a}_{-{\bm k}} \Big)
+ i\frac{a'}{a} \Big( -\hat{a}_{\bm k}\hat{a}_{-{\bm k}} 
+ \hat{a}_{-{\bm k}}^\dag \hat{a}_{\bm k}^\dag \Big) \Bigg] \, .
\end{equation}

Then, using the solutions \eqref{eq:rk-sol}, \eqref{eq:varphik-sol} and \eqref{eq:thetak-sol}, the evolution operator is factorized as
\begin{equation}
\label{eq:evolution-op}
\widehat{U}_{\bm k} = \widehat{R}_{\bm k} \widehat{S}_{\bm k} \, ,
\end{equation}
where
\begin{align}
\label{eq:rotation-op}
\widehat{R}_{\bm k} 
& = 
\exp \Big[ -i\theta_k 
\Big( 1 + \hat{a}_{\bm k}^\dag \hat{a}_{\bm k} 
+ \hat{a}_{-{\bm k}}^\dag \hat{a}_{-{\bm k}} \Big)
\Big] \, ,
\\
\label{eq:squeezing-op}
\widehat{S}_{\bm k} 
& = 
\exp \bigg[
\frac{r_k}{2} \Big( \hat{a}_{\bm k} \hat{a}_{-{\bm k}} e^{-2i\varphi_k} 
- \hat{a}_{-{\bm k}}^\dag \hat{a}_{\bm k}^\dag e^{2i\varphi_k} \Big) \bigg] \, .
\end{align}
$\widehat{R}_{\bm k}$ and $\widehat{S}_{\bm k}$ are called rotation and squeezing operators respectively.

Since the evolution operator \eqref{eq:evolution-op} is coming from the quadratic Hamiltonian, we can dump the whole evolution into the (initial vacuum) state in the Schr\"odinger picture. Since \eqref{eq:evolution-op} is factorized into \eqref{eq:rotation-op} and \eqref{eq:squeezing-op}, the action of each piece is also separated. For the rotation operator \eqref{eq:rotation-op}, as we expand the exponential, the operator terms annihilate the vacuum since it is first multiplied by the annihilation operators $\hat{a}_{\bm k}$ and $\hat{a}_{-{\bm k}}$. Thus only the first term, which originally accounts for the volume over which the field permeates, survives. That is, with the vacuum state defined in the standard manner as\footnote{This is also valid for our original operator $a_{\bm k}$, as the only difference is the volume normalization factor $L^{3/2}$.}
\begin{equation}
\label{eq:BDvacuum}
\hat{a}_{\bm k}|0\rangle_0 = 0 ~
\text{for all } {\bm k} \, ,
\end{equation}
the action of $\widehat{R}_{\bm k}$ on $|0\rangle_0$ is
\begin{equation}
\label{eq:rotation-action}
\widehat{R}_{\bm k}|0\rangle_0 
= e^{-i\theta_k} |0\rangle_0 \, .
\end{equation}
Thus the rotation operator $\widehat{R}_{\bm k}$ does on the initial vacuum state $|0\rangle_0$ nothing but producing an irrelevant phase $\theta_k$. This is natural, since $\widehat{R}_{\bm k}$ is coming from the decoupled, free part of the Hamiltonian $\calH_0$ for which the expansion of the universe is neglected, such that $\calH_0$ should do in the same manner as what is done in the Minkowski space. That is, the vacuum remains vacuum.

The squeezing operator \eqref{eq:squeezing-op}, we can rewrite in as
\begin{equation}
\widehat{S}_{\bm k} = \exp \bigg[ - e^{2i\varphi_k} \tanh 
\left( \frac{r_k}{2} \right) \hat{a}_{-{\bm k}}^\dag \hat{a}_{\bm k}^\dag \bigg]
\left[ \frac{1}{\cosh(r_k/2)} \right]^{\hat{a}_{\bm k}^\dag \hat{a}_{\bm k} 
+ a_{-{\bm k}}^\dag \hat{a}_{-{\bm k}} + 1}
\exp \bigg[ e^{-2i\varphi_k} \tanh \left( \frac{r_k}{2} \right) 
\hat{a}_{-{\bm k}} \hat{a}_{\bm k} \bigg] 
\, .
\end{equation}
Using $a^x = 1+ x\log{a} + x^2(\log{a})^2/2! + \cdots$ and again expanding the exponential, we find the operator term $\hat{a}_{-{\bm k}} \hat{a}_{\bm k}$ does nothing as it annihilates the vacuum, so that
\begin{equation}
\label{eq:squeezing-action}
\widehat{S}_{\bm k}|0\rangle_0 
= \frac{1}{\cosh(r_k/2)} \sum_{n=0}^\infty 
\bigg[ - e^{2i\varphi_k} \tanh \left( \frac{r_k}{2} \right) \bigg]^n
|n,{\bm k};n,-{\bm k} \rangle 
\, ,
\end{equation}
where
\begin{equation}
\label{eq:squeezedstate}
|n,{\bm k};n,-{\bm k} \rangle \equiv
\frac{1}{n!} \Big( \hat{a}_{-{\bm k}}^\dag \hat{a}_{\bm k}^\dag \Big)^n 
|0\rangle_0
\end{equation}
is the two-mode (with momenta ${\bm k}$ and $-{\bm k}$ for momentum conservation) state with the same occupation number $n$, i.e. $n$-particle state. Thus the squeezing operator $\widehat{S}_{\bm k}$, originated from the interacting part of the Hamiltonian $\calH_i$, is responsible for the creation of two quanta with momenta ${\bm k}$ and $-{\bm k}$. In other words, the cosmological vacuum fluctuations are amplified.

\section{Full reduced density matrix}
\label{app:fullrho}
\setcounter{equation}{0}

In this section we present the matrix for the Lindblad terms before any further consideration based on the configuration of momenta. The Lindblad terms for the evolution of the reduced density matrix may well be represented by a 7$\times$7 matrix, depending on how many operators are involved, running from 0 to 6 as:
\begin{equation}
\label{eq:matrix-schematic}
\frac{d\rho_\text{red}}{d\tau} =
-\frac{1}{2}
\begin{pmatrix}
\mathfrak{E}_{00} & 0 & \mathfrak{E}_{02} & 0 & \mathfrak{E}_{04} & 0 & 0
\\
0 & \mathfrak{E}_{11} & 0 & \mathfrak{E}_{13} & 0 & 0 & 0
\\
\mathfrak{E}_{20} & 0 & \mathfrak{E}_{22} & 0 & 0 & 0 & 0
\\
0 & \mathfrak{E}_{31} & 0 & 0 & 0 & 0 & 0 
\\
\mathfrak{E}_{40} & 0 & 0 & 0 & 0 & 0 & 0
\\
0 & 0 & 0 & 0 & 0 & 0 & 0
\\
0 & 0 & 0 & 0 & 0 & 0 & 0
\end{pmatrix}
+ h.c.
\, ,
\end{equation}
where each term is given as follows:

\begin{enumerate}

\item $\mathfrak{E}_{00}$: This is the simplest case where both the bra- and ket-basis vectors $\langle a|$ and $|b\rangle$ are the vacuum-evolved state. To obtain such terms, we multiply from the left $\big( U_{0,\calS}|0\rangle_\calS \big)^\dag = \langle0|_\calS U_{0,\calS}^\dag$ and from the right $U_{0,\calS}|0\rangle_\calS$. To have non-zero contributions, there must be the factor of the form $U_{0,\calS}|0\rangle_\calS \langle0|_\calS U^\dag_{0,\calS}$.

Two contributions are coming from the case, using the notations in \eqref{eq:Lindblad-terms}, $H = a^\dag_3a^\dag_2a_1$, $a^\dag_2a_1a^\dag_3$ and $a_1a^\dag_3a^\dag_2$ and $G = a_4a^\dag_6a^\dag_5$, $a^\dag_5a_4a^\dag_6$ and $a^\dag_6a^\dag_5a_4$. If either $H^\dag$ or $G$ has one self-contraction, 
\begin{align}
\label{eq:matrix00-2'}
\mathfrak{E}_{00}
& \supset
\frac{2}{(2\pi)^3} \delta^{(3)}({\bm q})
\int_\calS 
d^3k_1 
\int_\calE 
d^3k_4 
\sum_{\lambda_i}
\Big[ 2 h_1({\bm k}_1,{\bm q},-{\bm k}_1) + h_1({\bm k}_1,-{\bm k}_1,{\bm q}) \Big]
\nonumber\\
& \quad \times 
\Big[ 2 g_1^*({\bm k}_4,-{\bm q},-{\bm k}_4)
+ g_1^*({\bm k}_4,-{\bm k}_4,-{\bm q}) \Big]
+ \big( \calE \leftrightarrow \calS \big)
\, ,
\end{align}
which the product of one squeezed triangle with all the momenta in the system sector and another squeezed triangle with squeezed momenta in the environment sector, sharing a common momentum ${\bm q} \approx 0$: $(\calS\calS\calS)_\text{sq} \times (\calE\calE\calS)_\text{sq}$. Also note that the same momenta mean the same polarization indices, e.g. $h_1({\bm k}_1,{\bm q},-{\bm k}_1) g_1^*({\bm k}_4,-{\bm q},-{\bm k}_4)$ means $\lambda_1=\lambda_3$, $\lambda_2=\lambda_5$ and $\lambda_4=\lambda_6$. If both $H^\dag$ and $G$ have one self-contraction each at the same time, 
\begin{align}
\label{eq:matrix00-3'}
\mathfrak{E}_{00}
& \supset
\frac{2}{(2\pi)^3} \delta^{(3)}({\bm q}) 
\int_\calE 
d^3k_1
\int_\calE
d^3k_4
\sum_{\lambda_i}
\Big[ 2 h_1({\bm k}_1,{\bm q},-{\bm k}_1) + h_1({\bm k}_1,-{\bm k}_1,{\bm q}) \Big]
\nonumber\\
& \quad 
\times
\Big[ 2 g_1^*({\bm k}_4,-{\bm q},-{\bm k}_4)
+ g_1^*({\bm k}_4,-{\bm k}_4,-{\bm q}) \Big]
\, ,
\end{align}
which is a product of two squeezed triangles with two momenta in the environment sector respectively, sharing a common momentum ${\bm q} \approx 0$: $(\calE\calE\calS)_\text{sq} \times (\calE\calE\calS)_\text{sq}$.

The other two contributions are coming from the case $H = a^\dag_3a^\dag_2a^\dag_1$ and $G = a^\dag_6a^\dag_5a^\dag_4$. If there is one cross-contraction between $H^\dag$ and $G$,
\begin{align}
\label{eq:matrix00-5'}
\mathfrak{E}_{00}
& \supset
\frac{2}{(2\pi)^3} \delta^{(3)}({\bm q}) \sum_{\lambda_i}
\int_\calS d^3k_1 \int_\calE d^3k_2 
\nonumber\\
& \quad 
\times
\Big[ h_0({\bm k}_2,{\bm k}_1,-{\bm k}_{12}+{\bm q}) 
+ h_0({\bm k}_1,{\bm k}_2,-{\bm k}_{12}+{\bm q}) 
+ h_0({\bm k}_1,-{\bm k}_{12}+{\bm q},{\bm k}_2) \Big]
\nonumber\\
& \quad 
\times
\Big[
g_0^*(-{\bm k}_1,-{\bm k}_2,{\bm k}_{12}-{\bm q})
+ \big( \text{5 perm in arranging } -{\bm k}_1,-{\bm k}_2,{\bm k}_{12}-{\bm q} \text{ in } g_0^* \big)
\Big] 
\, , 
\end{align}
which is a product of two identical triangles with one momentum in the environment sector respectively, i.e. ${\bm k}_3 = -{\bm k}_{12} + {\bm q} \in {\bm k}_\calS$ with ${\bm q} \approx 0$: $(\calE\calS\calS) \times (\calE\calS\calS)$. If there are two cross-contractions between $H^\dag$ and $G$,
\begin{align}
\label{eq:matrix00-6'}
\mathfrak{E}_{00}
& \supset
\frac{2}{(2\pi)^3} \delta^{(3)}({\bm q}) \sum_{\lambda_i}
\int_\calS d^3k_1 \int_\calE d^3k_2 
\nonumber\\
& \quad 
\times
\Big[ h_0({\bm k}_2,{\bm k}_1,-{\bm k}_{12}+{\bm q}) 
+ h_0({\bm k}_1,{\bm k}_2,-{\bm k}_{12}+{\bm q}) 
+ h_0({\bm k}_1,-{\bm k}_{12}+{\bm q},{\bm k}_2) \Big]
\nonumber\\
& \quad 
\times
\Big[
g_0^*(-{\bm k}_1,-{\bm k}_2,{\bm k}_{12}-{\bm q})
+ \big( \text{5 perm in arranging } -{\bm k}_1,-{\bm k}_2,{\bm k}_{12}-{\bm q} \text{ in } g_0^* \big)
\Big] 
\, , 
\end{align}
which is a product of two identical triangles with two momenta in the environment sector respectively, i.e. ${\bm k}_3 = -{\bm k}_{12} + {\bm q} \in {\bm k}_\calE$: $(\calE\calE\calS) \times (\calE\calE\calS)$.

\item $\mathfrak{E}_{02}$: This corresponds to the case where the bra-basis vector $\langle a|$ is the vacuum-evolved state and the ket-basis vector $|b\rangle$ corresponds to the vacuum-evolved two-excited state with momenta ${\bm k}_a$ and ${\bm k}_b$. To obtain such terms, we multiply from the left $\big( U_{0,\calS}|0\rangle_\calS \big)^\dag = \langle0|_\calS U_{0,\calS}^\dag$ and from the right $U_{0,\calS}a^\dag_aa^\dag_b|0\rangle_\calS$. To have non-zero contributions, there must be the factor of the form $U_{0,\calS}|0\rangle_\calS \langle0|_\calS a_{a'}a_{b'} U_{0,\calS}^\dag$.

Two contributions are coming from the case $H = a^\dag_3a^\dag_2a_1$, $a^\dag_2a_1a^\dag_3$ and $a_1a^\dag_3a^\dag_2$ and $G = a_4^\dag a_5a_6$, $a_6a^\dag_4a_5$ and $a_5a_6a^\dag_4$. If either $H^\dag$ or $G$ has one self-contraction, 
\begin{align}
\label{eq:matrix02-2'}
\mathfrak{E}_{02}
& \supset
\delta^{(3)}({\bm q}_{a}) \delta^{(3)}({\bm q}_{b})
\bigg( \int_\calS d^3k_1 \int_\calE d^3k_4
+ \int_\calE d^3k_1 \int_\calS d^3k_4 \bigg)
\sum_{\lambda_i}
\nonumber\\
& \qquad 
\times
\Big[ 2 h_1({\bm k}_1,{\bm q}_a,-{\bm k}_1) + h_1({\bm k}_1,-{\bm k}_1,{\bm q}_a) \Big]
\Big[ 2 g_1({\bm k}_4,{\bm q}_b,-{\bm k}_4)
+ g_1({\bm k}_4,-{\bm k}_4,{\bm q}_b) \Big]
\nonumber\\
& \quad
+
3 \delta^{(3)}({\bm q}_{ab})
\int_\calE d^3k_1
\sum_{\lambda_i}
\Big[ 
2 h_1({\bm k}_1,{\bm q}_{ab},-{\bm k}_1) 
+
h_1({\bm k}_1,-{\bm k}_1,{\bm q}_{ab}) \Big] g_1(-{\bm q}_{ab},{\bm q}_a,{\bm q}_b)
\nonumber\\
& \quad
+
({\bm q}_a \leftrightarrow {\bm q}_b)
\, .
\end{align}
Here, in the first (second) four terms, the two momenta sets constitute independent squeezed triangles (${\bm q}_a \approx 0$ and ${\bm q}_b \approx 0$ but in general ${\bm q}_a \neq {\bm q}_b$), in one of which all in the system sector and the other two in the environment sector for ${\bm k}_1$ and ${\bm k}_4$ (${\bm k}_4$ and ${\bm k}_1$) respectively: $(\calS\calS\calS)_\text{sq} \times (\calE\calE\calS)_\text{sq}$. Meanwhile, the last two terms share the squeezed momentum denoted in common by ${\bm q}_{ab} \in {\bm k}_\calS$: $(\calE\calE\calS)_\text{sq} \times (\calE\calE\calS)_\text{sq}$. If both $H^\dag$ and $G$ have one self-contraction each at the same time, 
\begin{align}
\label{eq:matrix02-3'}
\mathfrak{E}_{02}
& \supset
\delta^{(3)}({\bm q}_{a}) \delta^{(3)}({\bm q}_{b})
\int_\calE d^3k_1 \int_\calE d^3k_4 
\sum_{\lambda_i}
\Big[ 2 h_1({\bm k}_1,{\bm q}_a,-{\bm k}_1) 
+ h_1({\bm k}_1,-{\bm k}_1,{\bm q}_a) \Big]
\nonumber\\
& \quad \times 
\Big[ 2 g_1({\bm k}_4,{\bm q}_b,-{\bm k}_4) 
+ g_1({\bm k}_4,-{\bm k}_4,{\bm q}_b) \Big]
+
({\bm q}_a \leftrightarrow {\bm q}_b)
\, ,
\end{align}
which is the product of two independent squeezed triangles with two momenta in the environment sector respectively: $(\calE\calE\calS)_\text{sq} \times (\calE\calE\calS)_\text{sq}$.

The other three contributions are coming from the case $H = a^\dag_3a^\dag_2a^\dag_1$ and $G = a_4a^\dag_6a^\dag_5$, $a^\dag_5a_4a^\dag_6$ and $a^\dag_6a^\dag_5a_4$. If there is one self-contraction in $G$,
\begin{align}
\label{eq:matrix02-7'}
\mathfrak{E}_{02}
& \supset
\delta^{(3)}({\bm q}_{ab})
\int_\calE
d^3k_4 
\sum_{\lambda_i}
\Big[ h_0(-{\bm q}_{ab},{\bm q}_a,{\bm q}_b) 
+ h_0({\bm q}_a,-{\bm q}_{ab},{\bm q}_b)
+ h_0({\bm q}_a,{\bm q}_b,-{\bm q}_{ab}) \Big]
\nonumber\\
& \quad 
\times
\Big[ g_1^*({\bm k}_4,-{\bm k}_4,{\bm q}_{ab})
+ g_1^*({\bm k}_4,{\bm q}_{ab},-{\bm k}_4) \Big]
+
({\bm q}_a \leftrightarrow {\bm q}_b)
\, ,
\end{align}
which is the product of two squeezed triangles with two momenta in the system sector in one triangle and in the environment sector in the other triangle, respectively, sharing ${\bm q}_{ab} \approx 0$ (and ${\bm q}_{ab} \in {\bm k}_\calS$): $(\calS\calS\calS)_\text{sq} \times (\calE\calE\calS)_\text{sq}$. If there is one cross-contraction between $H^\dag$ and $G$,
\begin{align}
\label{eq:matrix02-8'}
\mathfrak{E}_{02}
& \supset
\delta^{(3)}({\bm q}_{ab})
\int_\calE
d^3k_1 
\sum_{\lambda_i}
3 \Big[ h_0({\bm k}_1,{\bm q}_a,-{\bm k}_1+{\bm q}_b) 
+ h_0({\bm k}_1,-{\bm k}_1+{\bm q}_b,{\bm q}_a) 
\nonumber\\
& \hspace{10em}
+ h_0({\bm q}_a,{\bm k}_1,-{\bm k}_1+{\bm q}_b) 
+ h_0(-{\bm k}_1+{\bm q}_b,{\bm k}_1,{\bm q}_a) 
\nonumber\\
& \hspace{10em}
+ h_0({\bm q}_a,-{\bm k}_1+{\bm q}_b,{\bm k}_1) 
+ h_0(-{\bm k}_1+{\bm q}_b,{\bm q}_a,{\bm k}_1) \Big]
\nonumber\\
& \quad 
\times
\Big[ g_1^*({\bm q}_b,-{\bm k}_1,{\bm k}_1-{\bm q}_b)
+ g_1^*({\bm q}_b,{\bm k}_1-{\bm q}_b,-{\bm k}_1) \Big]
+
({\bm q}_a \leftrightarrow {\bm q}_b)
\, ,
\end{align}
which is the product of two identical triangles with one momentum being in the environment sector while the other two in the system sector and with ${\bm k}_1-{\bm q}_b \in {\bm k}_\calS$: $(\calE\calS\calS) \times (\calE\calS\calS)$. If there are two cross-contractions between $H^\dag$ and $G$,
\begin{align}
\label{eq:matrix02-10'}
\mathfrak{E}_{02}
& \supset
\delta^{(3)}({\bm q}_{ab})
\int_\calE
d^3k_1 
\sum_{\lambda_i}
3 \Big[ h_0({\bm k}_1,-{\bm k}_1+{\bm q}_b,{\bm q}_a) 
+ h_0({\bm k}_1,{\bm q}_a,-{\bm k}_1+{\bm q}_b) 
+ h_0({\bm q}_a,{\bm k}_1,-{\bm k}_1+{\bm q}_b) \Big]
\nonumber\\
& \quad 
\times
\Big[ g_1^*({\bm q}_b,-{\bm k}_1,{\bm k}_1-{\bm q}_b) 
+ g_1^*({\bm q}_b,{\bm k}_1-{\bm q}_b,-{\bm k}_1) \Big]
+
({\bm q}_a \leftrightarrow {\bm q}_b)
\, ,
\end{align}
which is the product of two triangles sharing one common momentum ${\bm k}_1$ in the environment sector with ${\bm k}_1-{\bm q}_b \in {\bm k}_\calE$: $(\calE\calE\calS) \times (\calE\calE\calS)$.

\item $\mathfrak{E}_{04}$: This corresponds to the case where the bra-basis vector $\langle a|$ is the vacuum-evolved state and the ket-basis vector $|b\rangle$ corresponds to the vacuum-evolved 4-excited state with momenta ${\bm k}_a$, ${\bm k}_b$, ${\bm k}_c$ and ${\bm k}_d$. To obtain such terms, we multiply from the left $\big( U_{0,\calS}|0\rangle_\calS \big)^\dag = \langle0|_\calS U_{0,\calS}^\dag$ and from the right $U_{0,\calS}a^\dag_aa^\dag_ba^\dag_ca^\dag_d|0\rangle_\calS$. To have non-zero contributions, there must be the factor of the form $U_{0,\calS}|0\rangle_\calS \langle0|_\calS a_{a'}a_{b'}a_{c'}a_{d'} U_{0,\calS}^\dag$.

One contribution is coming from the case $H = a^\dag_3a^\dag_2a_1$, $a^\dag_2a_1a^\dag_3$ and $a_1a^\dag_3a^\dag_2$ and $G = a_4a_5a_6$. Only one self-contraction in $H^\dag$ is possible to give
\begin{align}
\label{eq:matrix04-2'}
\mathfrak{E}_{04}
& \supset
2 \sum_{\lambda_i} (2\pi)^3 
\int_\calE
d^3k_1
\Big\{
\delta^{(3)}({\bm q}_a) \delta^{(3)}({\bm q}_{bcd})
\Big[
h_1({\bm k}_1,{\bm q}_a,-{\bm k}_1) 
+
h_1({\bm k}_1,-{\bm k}_1,{\bm q}_a) \Big] 
g_0({\bm q}_b,{\bm q}_c,{\bm q}_d) 
\nonumber\\
& \hspace{9em}
+ \big( \text{23 perm in arranging } {\bm q}_a,{\bm q}_b,{\bm q}_c,{\bm q}_d 
\text{ in } h_1 \text{ and } g_0 \big)
\Big\}
\, ,
\end{align}
which is the product of a squeezed triangle with two unsqueezed momenta in the environment sector and an arbitrary triangle in the system sector, not sharing any common momentum: $(\calE\calE\calS)_\text{sq} \times (\calS\calS\calS)$.

The other two contributions are coming from the case $H = a^\dag_3a^\dag_2a^\dag_1$ and $G = a^\dag_4a_5a_6$, $a_6a_4^\dag a_5$ and $a_5a_6a_4^\dag$. If $G$ has one self-contraction, 
\begin{align}
\label{eq:matrix04-4'}
\mathfrak{E}_{04}
& \supset
(2\pi)^3 
\int_\calE
d^3k_4
\sum_{\lambda_i} \delta^{(3)}({\bm q}_{abc}) \delta^{(3)}({\bm q}_d) 
h_0({\bm q}_a,{\bm q}_b,{\bm q}_c) 
\Big[ 2 g_1({\bm k}_4,{\bm q}_d,-{\bm k}_4)
+ g_1({\bm k}_4,-{\bm k}_4,{\bm q}_d) \Big]
\nonumber\\
& \quad
+ \big( \text{23 perm in arranging } {\bm q}_a,{\bm q}_b,{\bm q}_c,{\bm q}_d 
\text{ in } h_0 \text{ and } g_1 \big)
\, ,
\end{align}
which is the product of a squeezed triangle with two unsqueezed momenta in the environment sector and an arbitrary triangle in the system sector, not sharing any common momentum: $(\calE\calE\calS)_\text{sq} \times (\calS\calS\calS)$. If there is one cross-contraction between $H^\dag$ and $G$,
\begin{align}
\label{eq:matrix04-5'}
\mathfrak{E}_{04}
& \supset
3 (2\pi)^3 \sum_{\lambda_i} \delta^{(3)}({\bm q}_{abcd})
\Big[ h_0(-{\bm q}_{ab},{\bm q}_a,{\bm q}_b) 
+ h_0({\bm q}_a,-{\bm q}_{ab},{\bm q}_b) 
+ h_0({\bm q}_a,{\bm q}_b,-{\bm q}_{ab}) \Big]
g_1({\bm q}_{ab},{\bm q}_c,{\bm q}_d)
\nonumber\\
& \quad
+ \big( \text{23 perm in arranging } {\bm q}_a,{\bm q}_b,{\bm q}_c,{\bm q}_d 
\text{ in } h_0 \text{ and } g_1 \big)
\, ,
\end{align}
which is the product of two triangles sharing a common momentum ${\bm q}_{ab} \in {\bm k}_\calE$, yet the other momenta being in the system sector: $(\calE\calS\calS) \times (\calE\calS\calS)$.

\item $\mathfrak{E}_{11}$: This corresponds to the case where both the bra- and ket-basis vectors $\langle a|$ and $|b\rangle$ are the vacuum-evolved one-excited state. To obtain such terms, we multiply from the left $\big( U_{0,\calS}a^\dag_a|0\rangle_\calS \big)^\dag = \langle0|_\calS a_a U_{0,\calS}^\dag$ and from the right $U_{0,\calS}a^\dag_b|0\rangle_\calS$. To have non-zero contributions, there must be the factor of the form $U_{0,\calS}a^\dag_{a'}|0\rangle_\calS \langle0|_\calS a_{b'} U_{0,\calS}^\dag$.

Two contributions are coming from the case $H = a^\dag_3a^\dag_2a_1$, $a^\dag_2a_1a^\dag_3$ and $a_1a^\dag_3a^\dag_2$ and $G = a_4a^\dag_6a^\dag_5$, $a^\dag_5a_4a^\dag_6$ and $a^\dag_6a^\dag_5a_4$. If either $H^\dag$ or $G$ has one self-contraction, 
\begin{align}
\label{eq:matrix11-2'}
\mathfrak{E}_{11}
& \supset
- 2 \delta^{(3)}({\bm q}_{a}) \delta^{(3)}({\bm q}_{b})
\bigg( 
\int_\calS d^3k_1 
\int_\calE d^3k_4 
+
\int_\calE d^3k_1
\int_\calS d^3k_4
\bigg)
\sum_{\lambda_i}
\nonumber\\
& \quad \times
\Big[ 2 h_1({\bm k}_1,{\bm q}_a,-{\bm k}_1) 
+
h_1({\bm k}_1,-{\bm k}_1,{\bm q}_a) \Big]
\Big[ g_1^*({\bm k}_4,-{\bm k}_4,{\bm q}_b) + 2 g_1^*({\bm k}_4,{\bm q}_b,-{\bm k}_4) \Big]
\, .
\end{align}
This is the product of a squeezed triangle with all momenta in the system sector and another, independent squeezed triangle with two momenta in the environment sector: $(\calS\calS\calS)_\text{sq} \times (\calE\calE\calS)_\text{sq}$. If both $H^\dag$ and $G$ have one self-contraction each at the same time,
\begin{align}
\label{eq:matrix11-3'}
\mathfrak{E}_{11}
& \supset
- 2 \delta^{(3)}({\bm q}_{a}) \delta^{(3)}({\bm q}_{b})
\int_\calE
d^3k_1 
\int_\calE
d^3k_4
\sum_{\lambda_i}
\nonumber\\
& \quad \times
\Big[ 2 h_1({\bm k}_1,{\bm q}_a,-{\bm k}_1) 
+ h_1({\bm k}_1,-{\bm k}_1,{\bm q}_a) \Big]
\Big[ g_1^*({\bm k}_4,-{\bm k}_4,{\bm q}_b)
+
2 g_1^*({\bm k}_4,{\bm q}_b,-{\bm k}_4) \Big]
\, ,
\end{align}
which is the product of two independent squeezed triangles with two momenta in the environment sector respectively: $(\calE\calE\calS)_\text{sq} \times (\calE\calE\calS)_\text{sq}$.

The other one contribution is coming from the case $H = a^\dag_3a^\dag_2a^\dag_1$ and $G = a^\dag_6a^\dag_5a^\dag_4$. If there are two cross-contractions between $H^\dag$ and $G$,
\begin{align}
\label{eq:matrix11-4'}
\mathfrak{E}_{11}
& \supset
- 2 \delta^{(3)}({\bm q}_{ab})
\int_\calE
d^3k_1 
\sum_{\lambda_i}
\Big[
h_0({\bm q}_a,{\bm k}_1,-{\bm k}_1+{\bm q}_b) 
+ h_0({\bm k}_1,{\bm q}_a,-{\bm k}_1+{\bm q}_b) 
+ h_0({\bm k}_1,-{\bm k}_1+{\bm q}_b,{\bm q}_a) \Big]
\nonumber\\
& \quad \times
\Big[ g_0^*({\bm q}_b,-{\bm k}_1,{\bm k}_1-{\bm q}_b)
+ \big( \text{5 perm in arranging } {\bm q}_b,-{\bm k}_1,{\bm k}_1-{\bm q}_b
\text{ in } g_0^* \big)
\Big]
\, .
\end{align}
This is the product of 2 identical (because $q_a = q_b$ due to the delta function) triangles with ${\bm k}_1 \in {\bm k}_\calE$ and ${\bm k}_1-{\bm q}_b \in {\bm k}_\calE$: $(\calE\calE\calS)_\text{sq} \times (\calE\calE\calS)_\text{sq}$.

\item $\mathfrak{E}_{13}$: This corresponds to the case where the bra-basis vector $\langle a|$ is the vacuum-evolved single-excited state and the ket-basis vector $|b\rangle$ corresponds to the vacuum-evolved 3-excited state. To obtain such terms, we multiply from the left $\big( U_{0,\calS}a^\dag_a|0\rangle_\calS \big)^\dag = \langle0|_\calS a_a U_{0,\calS}^\dag$ and from the right $U_{0,\calS}a^\dag_ba^\dag_ca^\dag_d|0\rangle_\calS$. To have non-zero contributions, there must be the factor of the form $U_{0,\calS}a^\dag_{a'}|0\rangle_\calS \langle0|_\calS a_{b'}a_{c'}a_{d'} U_{0,\calS}^\dag$.

One contribution is coming from the case $H = a^\dag_3a^\dag_2a^\dag_1$ and $G = a_4a^\dag_6a^\dag_5$, $a^\dag_5a_4a^\dag_6$ and $a^\dag_6a^\dag_5a_4$. Only one self-contraction in $G$ is possible to give
\begin{align}
\label{eq:matrix13-1'}
\mathfrak{E}_{13}
& =
-2 (2\pi)^{3\cdot2} \delta^{(3)}({\bm q}_{abc}) \delta^{(3)}({\bm q}_{d})
\int_\calE
d^3k_4
\sum_{\lambda_i}
h_0({\bm q}_a,{\bm q}_b,{\bm q}_c) 
\Big[ g_1^*(-{\bm k}_4,{\bm k}_4,{\bm q}_d)
+ 2 g_1^*(-{\bm k}_4,{\bm q}_d,{\bm k}_4) \Big]
\nonumber\\
& \quad
+ \big( \text{5 perm in arranging } {\bm q}_a,{\bm q}_b,{\bm q}_c
\text{ in } h_0 \big)
\, ,
\end{align}
which is the product of one squeezed triangle with two momenta in the environment sector and another independent triangle with an arbitrary shape in the system sector: $(\calE\calE\calS)_\text{sq} \times (\calS\calS\calS)$.

\item $\mathfrak{E}_{22}$: This corresponds to the case where both the bra- and ket-basis vectors $\langle a|$ and $|b\rangle$ are the vacuum-evolved two-excited states. To obtain such terms, we multiply from the left $\big( U_{0,\calS}a^\dag_aa^\dag_b|0\rangle_\calS \big)^\dag = \langle0|_\calS a_ba_a U_{0,\calS}^\dag$ and from the right $U_{0,\calS}a^\dag_ca^\dag_d|0\rangle_\calS$. To have non-zero contributions, there must be the factor of the form $U_{0,\calS}a^\dag_{a'}a^\dag_{b'}|0\rangle_\calS \langle0|_\calS a_{c'}a_{d'} U_{0,\calS}^\dag$.

One contribution is coming from the case $H = a^\dag_3a^\dag_2a^\dag_1$ and $G = a^\dag_6a^\dag_5a^\dag_4$. If there is one cross-contraction between $H^\dag$ and $G$,
\begin{align}
\label{eq:matrix22-1'}
\mathfrak{E}_{22}
& =
- 2 \delta^{(3)}({\bm q}_{abcd})
\sum_{\lambda_i}
\Big[ h_0({\bm q}_{cd},{\bm q}_a,{\bm q}_b) 
+ h_0({\bm q}_a,{\bm q}_{cd},{\bm q}_b) 
+ h_0({\bm q}_a,{\bm q}_b,{\bm q}_{cd}) \Big]
\nonumber\\
& \qquad \times
\Big[ g_0^*(-{\bm q}_{cd},{\bm q}_c,{\bm q}_d)
+ g_0^*({\bm q}_c,-{\bm q}_{cd},{\bm q}_d)
+ g_0^*({\bm q}_c,{\bm q}_d,-{\bm q}_{cd}) \Big]
\nonumber\\
& \quad
+ \big( \text{3 perm in arranging } {\bm q}_a, {\bm q}_b \text{ in } h_0
\text{ and } {\bm q}_c, {\bm q}_d \text{ in } g_0^* \big)
\, ,
\end{align}
where ${\bm q}_{ab} = -{\bm q}_{cd} \in {\bm k}_\calE$. This is the product of two triangles sharing a common momentum ${\bm q}_{ab} = -{\bm q}_{cd}$ in the environment sector, while the rest momenta in the system sector: $(\calE\calS\calS) \times (\calE\calS\calS)$.

\end{enumerate}

Thus, to summarize, the non-zero elements of $d\rho_\text{red}/d\tau$ in the basis of squeezed states given by \eqref{eq:matrix-schematic} are
\begin{align}
\mathfrak{E}_{00} & = 
\eqref{eq:matrix00-2'} + \eqref{eq:matrix00-3'}
+ \eqref{eq:matrix00-5'} + \eqref{eq:matrix00-6'}
\, ,
\\
\mathfrak{E}_{02} & = 
\eqref{eq:matrix02-2'} + \eqref{eq:matrix02-3'}
+ \eqref{eq:matrix02-7'} + \eqref{eq:matrix02-8'} 
+ \eqref{eq:matrix02-10'}
\, ,
\\
\mathfrak{E}_{04} & = \eqref{eq:matrix04-2'}
+ \eqref{eq:matrix04-4'} + \eqref{eq:matrix04-5'}
\, ,
\\
\mathfrak{E}_{11} & = \eqref{eq:matrix11-2'} 
+ \eqref{eq:matrix11-3'} + \eqref{eq:matrix11-4'}
\, ,
\\
\mathfrak{E}_{13} & = \eqref{eq:matrix13-1'}
\, ,
\\
\mathfrak{E}_{22} & = \eqref{eq:matrix22-1'}
\, .
\end{align}
The other non-zero elements in \eqref{eq:matrix-schematic} not explicitly presented, $\mathfrak{E}_{20}$, $\mathfrak{E}_{31}$ and $\mathfrak{E}_{40}$, are the counterparts of $\mathfrak{E}_{02}$, $\mathfrak{E}_{13}$ and $\mathfrak{E}_{04}$ respectively, and can be computed almost identically.

\section{Reduction of density matrix elements}
\label{app:reduction}
\setcounter{equation}{0}

In the previous section, many configurations we have found are not arbitrary but specific, e.g. squeezed. This would align the momenta in a specific manner to simplify the polarization tensors. For explicit check, we adopt the unit vectors in the spherical coordinate system in such a way that the radial unit vector denotes the direction $\hat{\bm k}$, and the polar unit vector $\hat{\bm e}_1$ and the azimuthal unit vector $\hat{\bm e}_2$ denote the orientation in terms of the polar and azimuthal angles $\theta$ and $\psi$ orthogonal to $\hat{\bm k}$:
\begin{equation}
\begin{split}
\hat{\bm k} & = (\sin\theta\cos\psi,\sin\theta\sin\psi,\cos\theta) \, ,
\\
\hat{\bm e}_1 & = (\cos\theta\cos\psi,\cos\theta\sin\psi,-\sin\theta) \, ,
\\
\hat{\bm e}_2 & = (-\sin\psi,\cos\psi,0) \, .
\end{split}
\end{equation}
Then the polarization tensors $e_{ij}^+$ and $e_{ij}^\times$ can be constructed by $\hat{\bm e}_1$ and $\hat{\bm e}_2$ as
\begin{align}
\label{eq:ptensor-p}
e_{ij}^+ 
& =
\big( \hat{e}_1 \big)_i \big( \hat{e}_1 \big)_j - \big( \hat{e}_2 \big)_i \big( \hat{e}_2 \big)_j
=
\begin{pmatrix}
\cos^2\theta\cos^2\psi - \sin^2\psi & (1+\cos^2\theta)\sin\psi\cos\psi & -\sin\theta\cos\theta\cos\psi
\\
(1+\cos^2\theta)\sin\psi\cos\psi & \cos^2\theta\sin^2\psi - \cos^2\psi & -\sin\theta\cos\theta\sin\psi
\\
-\sin\theta\cos\theta\cos\psi & -\sin\theta\cos\theta\sin\psi & \sin^2\theta
\end{pmatrix}
\, ,
\\
\label{eq:ptensor-c}
e_{ij}^\times
& =
\big( \hat{e}_1 \big)_i \big( \hat{e}_2 \big)_j + \big( \hat{e}_2 \big)_i \big( \hat{e}_1 \big)_j
=
\begin{pmatrix}
-2\cos\theta\sin\psi\cos\psi & \cos\theta(\cos^2\psi - \sin^2\psi) & \sin\theta\sin\psi
\\
\cos\theta(\cos^2\psi - \sin^2\psi) & 2\cos\theta\sin\psi\cos\psi & -\sin\theta\cos\psi
\\
\sin\theta\sin\psi & -\sin\theta\cos\psi & 0
\end{pmatrix}
\, .
\end{align}
To check, let us consider the well-known case that ${\bm k}$ is aligned along $z$-direction. This corresponds to $\theta = \psi = 0$ so that $\hat{\bm k} = (0,0,1)$. Then the polarization tensors become
\begin{equation}
\label{eq:kz-ptensor}
e_{ij}^+ =
\begin{pmatrix}
1 & 0 & 0 \\ 0 & -1 & 0 \\ 0 & 0 & 0
\end{pmatrix}
\quad \text{and} \quad
e_{ij}^\times =
\begin{pmatrix}
0 & 1 & 0 \\ 1 & 0 & 0 \\ 0 & 0 & 0
\end{pmatrix}
\, ,
\end{equation}
reproducing the well-known results.

Then it is straightforward to compute all the possible combinations of the polarization matrices. For example, in \eqref{eq:matrix00-2'}, we can see from \eqref{eq:coeff-c1}, \eqref{eq:coeff-c2} and \eqref{eq:coeff-c3}, we need to consider the combinations of the polarization indices
\begin{equation}
\sum_{\lambda_i} \big( \lambda_1\lambda_1\lambda_q \big) \big( \lambda_4\lambda_4\lambda_q )
=
(+++)(+++) + (++\times)(++\times) + \cdots
\, .
\end{equation}
By making use of the explicit form of the polarization matrices \eqref{eq:ptensor-p} and \eqref{eq:ptensor-c}, we can see that all terms in $d\rho_\text{red}/d\tau$ given by \eqref{eq:matrix-schematic} vanish, except for \eqref{eq:matrix-00}, \eqref{eq:matrix-02}, \eqref{eq:matrix-11} and \eqref{eq:matrix-20}. Let us here explicitly consider the configuration $\calE\calS\calS$, which corresponds to the combination of one environment mode and two system modes. Such terms include \eqref{eq:matrix00-5'}, \eqref{eq:matrix02-8'}, \eqref{eq:matrix04-5'} and \eqref{eq:matrix22-1'} (also one contribution in $\mathfrak{E}_{20}$ and $\mathfrak{E}_{40}$ each). Being in the system and environment sector, we demand that the corresponding mode should be smaller and greater than the horizon scale $\calH$. To form a triangle, we also demand that the sum of the amplitudes of the two system modes be greater than that of one environment mode. So, if $k_1$ and $k_2$ are in the system sector and $k_3$ is in the environment sector, the following relation should holds:
\begin{equation}
k_1, k_2 < \calH < k_3 = \sqrt{k_1^2+k_2^2 + 2k_1k_2\cos\theta_{12}} < 2\calH \, ,
\end{equation}
where $\theta_{12}$ is the angle between ${\bm k}_1$ and ${\bm k}_2$. To saturate the bounds, $\theta_{12}$ should lie between 0 (for $k_3 = 2\calH$ with $k_1=k_2=\calH$) and $2\pi/3$ (for $k_3 = \calH$ with $k_1=k_2=\calH$). Thus, in this case the distinction between system and environment modes is not very clear, in the sense that they are different within only a factor of 2 or so. Thus, to make the difference between system and environment modes as prominent as possible to keep the validity of our effective approach of distinguishing system and environment, we take the extreme case of the folded configuration $\theta_{12}=0$ with $({\bm k}_1,{\bm k}_2,{\bm k}_3) = ({\bm k},{\bm k},-2{\bm k})$, i.e. ${\bm k}_3$ is the environment mode. Then, all polarization tensors are transverse to all momenta: $e^a_{ij}k_b^i = 0$ for $a,b = 1,2,3$. This is obvious if we write, say, ${\bm k}_1 = {\bm k}_2 = (0,0,k)$ and ${\bm k}_3 = (0,0,-2k)$, and multiply them to $e_{ij}^+$ and $e_{ij}^\times$ in \eqref{eq:kz-ptensor}. Then we are only left with the terms that contain the trace of the product of three polarization tensors, $e_{ij}^{\lambda_a}({\bm k}_a)e_{jk}^{\lambda_b}({\bm k}_b)e_{ki}^{\lambda_c}({\bm k}_c)$. Further, for $\hat{\bm k} = (0,0,-1)$, $\theta=\pi$ and $\psi = 0$ so that from \eqref{eq:ptensor-p} and \eqref{eq:ptensor-c} we can explicitly find $e_{ij}^+$ is the same as that in \eqref{eq:kz-ptensor} but $e_{ij}^\times$ has the opposite sign to that in \eqref{eq:kz-ptensor}. Thus, for all 8 combination of polarizations, the individual trace for each polarization vanishes:
\begin{equation}
\label{eq:ESS-polproduct}
{\rm Tr} \Big[ e^+(\hat{\bm k}_1)e^+(\hat{\bm k}_2)e^+(\hat{\bm k}_3) \Big]
=
{\rm Tr} \Big[ e^\times(\hat{\bm k}_1)e^+(\hat{\bm k}_2)e^+(\hat{\bm k}_3) \Big]
=
\cdots
= 0
\, .
\end{equation}
Thus, in the folded configuration which should be most significant for $\calE\calS\calS$, $c_1 = c_{2i} = c_{3i} = 0$ so that $h_0 = h_1 = g_0 = g_1 = 0$.

\section{Direct computation of reduced density matrix}
\label{app:matrix}
\setcounter{equation}{0}

We consider in the same squeezed basis \eqref{eq:matrixbases} the reduced density matrix given by \eqref{eq:rho_red}, where the full density matrix before reduction $\rho(\tau)$ is given by \eqref{eq:sol-s.pic}. With $\rho(\tau_0) = |0\rangle_\calE|0\rangle_\calS\langle0|_\calS|\langle0|_\calE$ and $U_0 = U_{0,\calS}U_{0,\calE}$, we find
\begin{align}
\label{eq:rho_red-explicit}
\rho_\text{red}(\tau)
& =
\sum_i \Big\langle \calE_i \Big| U_0(\tau,\tau_0) \rho(\tau_0) 
U_0^\dag(\tau,\tau_0) \Big| \calE_i \Big\rangle
\nonumber\\
& 
\quad
- i \sum_i \Big\langle \calE_i \Big| U_0(\tau,\tau_0)
\int_{\tau_0}^\tau d\tau_1 \big[ H_{\text{int},I}(\tau_1), \rho(\tau_0) \big] 
U_0^\dag(\tau,\tau_0) \Big| \calE_i \Big\rangle
\nonumber\\
&
\quad
+ (-i)^2 \sum_i \Big\langle \calE_i \Big| U_0(\tau,\tau_0)
\int_{\tau_0}^\tau d\tau_2 \int_{\tau_0}^{\tau_2} d\tau_1 
\big[ H_{\text{int},I}(\tau_2), \big[ H_{\text{int},I}(\tau_1), \rho(\tau_0) \big] \big]
U_0^\dag(\tau,\tau_0) \Big| \calE_i \Big\rangle
+ \cdots
\, .
\end{align}
Note that the first term is just trivial: since at free level the system and environment sectors commute,
\begin{align}
\sum_i \Big\langle \calE_i \Big| U_0(\tau,\tau_0) \rho(\tau_0) 
U_0^\dag(\tau,\tau_0) \Big| \calE_i \Big\rangle
=
U_{0,\calS}|0\rangle_\calS\langle0|_\calS U^\dag_{0,\calS}
= 
\rho_\text{red}^{(0)}(\tau)
\, .
\end{align}
As mentioned in \eqref{eq:rho0red-initial}, this is the free-evolved density matrix of the system states $\rho_\calS(\tau)$.

Now we turn to find the matrix element of $\rho_\text{red}(\tau)$ in the squeezed state basis \eqref{eq:matrixbases}. For the first term of \eqref{eq:rho_red-explicit}, only the $U_{0,\calS}|0\rangle_\calS$-row and $U_{0,\calS}|0\rangle_\calS$-column is non-zero and is given by
\begin{align}
&
\langle0|_\calS U_{0,\calS}^\dag \times
\text{1st term of \eqref{eq:rho_red-explicit}}
\times U_{0,\calS}|0\rangle_\calS
=
\sum_i \big\langle 0 \big|_\calS U_{0,\calS}^\dag \big\langle \calE_i \big| 
U_0(\tau,\tau_0) \rho(\tau_0) U_0^\dag(\tau,\tau_0) \big| \calE_i \big\rangle
U_{0,\calS} \big| 0 \big\rangle_\calS
=
1 
\end{align}
with our choice $\sum_i \big\langle\calE_i| U_{0,\calE} \big|\calE\big\rangle = 1$, and other components all vanishes.

For the second term of \eqref{eq:rho_red-explicit}, let us first write $H_{\text{int},I}$ explicitly as \eqref{eq:Hint_UV-IR2}. Since $H_{\text{int},I}$ is Hermitian, we can use $H_{\text{int},I} = H^\dag_{\text{int},I}$ for the second term, which will help further calculations as before. This gives
\begin{align}
\text{2nd term of \eqref{eq:rho_red-explicit}}
& =
- i \sum_i \Big[ \big\langle \calE_i \big| U_0 g^{mn}(\tau)G_{\calE,m}(\tau_0)G_{\calS,n}(\tau_0)
\big|0\big\rangle_\calE \big|0\big\rangle_\calS \big\langle0\big|_\calS \big\langle0\big|_\calE
U_0^\dag \big| \calE_i \big\rangle
\nonumber\\
&
\hspace{5em}
- \big\langle \calE_i \big| U_0 
\big|0\big\rangle_\calE \big|0\big\rangle_\calS \big\langle0\big|_\calS \big\langle0\big|_\calE
g^{pq*}(\tau)G^\dag_{\calE,p}(\tau_0)G^\dag_{\calS,q}(\tau_0)
U_0^\dag \big| \calE_i \big\rangle
\Big]
\nonumber\\
& =
- i \Big[
g^{mn}(\tau) \big\langle0\big|_\calE G_{\calE,m}(\tau_0) \big|0\big\rangle_\calE
U_{0,\calS} G_{\calS,n}(\tau_0) \big|0\big\rangle_\calS \big\langle0\big|_\calS U_{0,\calS}^\dag 
\nonumber\\
& 
\hspace{3em}
- g^{pq*}(\tau)
U_{0,\calS} \big|0\big\rangle_\calS \big\langle0\big|_\calS G^\dag_{\calS,q}(\tau_0) U_{0,\calS}^\dag
\big\langle0\big|_\calE G^\dag_{\calE,p}(\tau_0) \big|0\big\rangle_\calE 
\Big]
\, .
\end{align}
Since the interaction Hamiltonian contains three combinations of (initial) creation and annihilation operators, $\big\langle0\big|_\calE G_{\calE,m}(\tau_0) \big|0\big\rangle_\calE$ and $\big\langle0\big|_\calE G^\dag_{\calE,p}(\tau_0) \big|0\big\rangle_\calE$ identically vanish. Thus, there is no surviving contribution from the second term of \eqref{eq:rho_red-explicit}.

For the third term of \eqref{eq:rho_red-explicit}, since there are two interaction Hamiltonians, we use the Hermiticity of $H_{\text{int},I}$ in such a way that the third term of \eqref{eq:rho_red-explicit} resembles the Lindblad terms in \eqref{eq:Lindblad}. That is, for simplicity denoting $\int_{\tau_0}^\tau d\tau_1 H_{\text{int},I}(\tau_1) \equiv H_{\text{int},I}(\tau)$, we find
\begin{align}
\label{eq:rho_red-3rd}
\text{3rd term of \eqref{eq:rho_red-explicit}}
& = 
\sum_i \Big\langle \calE_i \Big| U_0 \Big[
H_{\text{int},I}(\tau) H_{\text{int},I}(\tau_2) \rho_0
+ \rho_0 H_{\text{int},I}(\tau_2) H_{\text{int},I}(\tau)
\nonumber\\
& 
\hspace{6em}
- H_{\text{int},I}(\tau) \rho_0 H_{\text{int},I}(\tau_2) 
- H_{\text{int},I}(\tau_2) \rho_0 H_{\text{int},I}(\tau)
\Big] U_0^\dag \Big| \calE_i \Big\rangle
\nonumber\\
& =
\sum_i \Big\langle \calE_i \Big| U_0 \Big[
H_{\text{int},I}^\dag(\tau) H_{\text{int},I}(\tau_2) \rho_0
+ \rho_0 H_{\text{int},I}^\dag(\tau_2) H_{\text{int},I}(\tau)
\nonumber\\
& 
\hspace{6em}
- H_{\text{int},I}(\tau) \rho_0 H_{\text{int},I}^\dag(\tau_2) 
- H_{\text{int},I}(\tau_2) \rho_0 H_{\text{int},I}^\dag(\tau)
\Big] U_0^\dag \Big| \calE_i \Big\rangle
\, .
\end{align}
Now using \eqref{eq:Hint_UV-IR2}, we find
\begin{align}
\label{eq:rho_red-3rd2}
\eqref{eq:rho_red-3rd}
& =
- \Bigg\{
g^{mn*}(\tau) g^{pq}(\tau_2) 
\big\langle0\big|_\calE G_{\calE,m}^\dag(\tau_0) G_{\calE,p}(\tau_0) \big|0\big\rangle_\calE
\nonumber\\
&
\hspace{3em}
\times
\bigg[
U_{0,\calS} G_{\calS,n}^\dag(\tau_0) G_{\calS,q}(\tau_0) 
\big|0\big\rangle_\calS \big\langle0\big|_\calS U_{0,\calS}^\dag
-
U_{0,\calS} G_{\calS,q}(\tau_0) \big|0\big\rangle_\calS
\big\langle0\big|_\calS G_{\calS,n}^\dag(\tau_0) U_{0,\calS}^\dag
\bigg]
+ h.c.
\Bigg\}
\, .
\end{align}
We see that the structure of \eqref{eq:rho_red-3rd2} is very similar to \eqref{eq:Lindblad-terms}. Three differences are 1) the time- and momentum-dependent coefficient is the product of two $g(\tau)$'s, 2) there is no prefactor of 1/2, and 3) the terms we have encountered during the calculation of $d\rho_\text{red}/d\tau$ in which $\rho_\text{red}$ is on the leftmost do not appear but they appear as the Hermitian conjugate terms. Then, we find $\rho_\text{red}$ as:
\begin{equation}
\label{eq:rhored-matrix-schematic2}
\rho_\text{red} =
\begin{pmatrix}
1 & 0_{1\times6}
\\
0_{6\times1} & 0_{6\times6}
\end{pmatrix}
-
\begin{pmatrix}
\widetilde{\mathfrak{E}}_{00} & 0 & 0 & 0 & 0 & 0 & 0
\\
0 & \widetilde{\mathfrak{E}}_{11} & 0 & 0 & 0 & 0 & 0
\\
\widetilde{\mathfrak{E}}_{20} & 0 & 0 & 0 & 0 & 0 & 0
\\
0 & 0 & 0 & 0 & 0 & 0 & 0 
\\
0 & 0 & 0 & 0 & 0 & 0 & 0
\\
0 & 0 & 0 & 0 & 0 & 0 & 0
\\
0 & 0 & 0 & 0 & 0 & 0 & 0
\end{pmatrix}
+ h.c.
\, ,
\end{equation}
where 
\begin{align}
\widetilde{\mathfrak{E}}_{00} + \widetilde{\mathfrak{E}}_{00}^*
& =
\frac{1}{(2\pi)^3} \delta^{(3)}({\bm q}) 
\int_\calS d^3k_1 \int_\calE d^3k_2 \sum_{\lambda_i}
g_0({\bm k}_1,{\bm k}_2,-{\bm k}_{12}+{\bm q};\tau)
\times 6g_0^*(-{\bm k}_1,-{\bm k}_2,{\bm k}_{12}-{\bm q};\tau_2)
+ c.c.
\nonumber\\
& =
\frac{2}{(2\pi)^3} \delta^{(3)}({\bm q}) 
\frac{-H^2}{\mpl^2} \frac{4}{9\tau^3}
8\pi^2 \calC_{\calS\calE} \, ,
\\
\widetilde{\mathfrak{E}}_{11} + \widetilde{\mathfrak{E}}_{11}^*
& =
- \delta^{(3)}({\bm q}_{ab}) \int_\calE d^3k_1 \sum_{\lambda_i}
3 g_0({\bm q},{\bm k}_1,-{\bm k}_1-{\bm q};\tau) \times 6 g_0^*(-{\bm q},-{\bm k}_1,{\bm k}_1+{\bm q};\tau_2)
+ c.c.
\nonumber\\
& =
36 \delta^{(3)}({\bm q}_{ab}) \delta_{\lambda_a \lambda_b}
\frac{H^2}{\mpl^2} \frac{4}{9\tau^3}
\frac{2\pi}{q^3} \calC_\calE 
\, ,
\\
\widetilde{\mathfrak{E}}_{20}
& =
3 \delta^{(3)}({\bm q}_{ab}) \int_\calE d^3k_1 \sum_{\lambda_i}
g_1({\bm q},{\bm k}_1,-{\bm k}_1-{\bm q};\tau)
\times 6 g_0^*(-{\bm q},-{\bm k}_1,{\bm k}_1+{\bm q};\tau_2) \times 2
\nonumber\\
& =
36 \delta^{(3)}({\bm q}_{ab}) \delta_{\lambda_a \lambda_b}
\frac{H^2}{\mpl^2} \frac{4}{9\tau^3} \frac{2\pi}{q^3} 
e^{2iq\tau} \calC_\calE
=
\widetilde{\mathfrak{E}}_{02}^*
\, .
\end{align}
Thus, except for the initial value $\mathfrak{E}_{00}=1$, each element of $\rho_\text{red}$ is precisely the time integral of the corresponding element in $d\rho_\text{red}/d\tau$ in the limit $\tau \to 0$ as it should be.

\end{document}